\documentclass[sigconf,nonacm]{acmart}

\usepackage{booktabs} 
\usepackage{multirow}
\usepackage{graphicx}
\AtBeginDocument{%
  }

\begin{document}

\title{WADEPre: A Wavelet-based Decomposition Model for Extreme Precipitation Nowcasting with Multi-Scale Learning}

\author{Baitian Liu}
\email{sonderliu@hdu.edu.cn}
\orcid{0009-0003-5770-4770}
\affiliation{
  \institution{Hangzhou Dianzi University}
  \city{Hangzhou}
  \state{Zhejiang}
  \country{China}
}

\author{Haiping Zhang}
\email{zhanghp@hdu.edu.cn}
\orcid{0009-0002-0038-0771}
\affiliation{
  \institution{Hangzhou Dianzi University}
  \city{Hangzhou}
  \state{Zhejiang}
  \country{China}
}

\author{Huiling Yuan}
\email{yuanhl@nju.edu.cn}
\orcid{0000-0003-4725-9039}
\affiliation{
  \institution{Nanjing University}
  \city{Nanjing}
  \state{Jiangsu}
  \country{China}
}

\author{Dongjing Wang}
\email{dongjing.wang@hdu.edu.cn}
\orcid{0000-0003-2152-0446}
\affiliation{
  \institution{Hangzhou Dianzi University}
  \city{Hangzhou}
  \state{Zhejiang}
  \country{China}
}

\author{Ying Li}
\orcid{0009-0005-7428-1479}
\email{ying.li@mail.bnu.edu.cn}
\author{Feng Chen}
\email{chenfeng@mail.iap.ac.cn}
\orcid{0000-0002-0105-3273}
\affiliation{
  \institution{Zhejiang Institute of Meteorological Sciences}
  \city{Hangzhou}
  \state{Zhejiang}
  \country{China}
}

%\affiliation{
%  \institution{Zhejiang Institute of Meteorological Sciences}
%  \city{Hangzhou}
%  \state{Zhejiang}
%  \country{China}
%}
%

\author{Hao Wu}
\email{wuhao@hdu.edu.cn}
\orcid{0009-0003-0360-4942}
\authornote{Corresponding author.}
\affiliation{
  \institution{Hangzhou Dianzi University}
  \city{Hangzhou}
  \state{Zhejiang}
  \country{China}
}

\renewcommand{\shortauthors}{Liu et al.}

\begin{abstract}
The heavy-tailed nature of precipitation intensity impedes precise precipitation nowcasting. Standard models that optimize pixel-wise losses are prone to regression-to-the-mean bias, which blurs extreme values. Existing Fourier-based methods also lack the spatial localization needed to resolve transient convective cells. To overcome these intrinsic limitations, we propose WADEPre, a wavelet-based decomposition model for extreme precipitation that transitions the modeling into the wavelet domain. By leveraging the Discrete Wavelet Transform for explicit decomposition, WADEPre employs a dual-branch architecture: an Approximation Network to model stable, low-frequency advection, isolating deterministic trends from statistical bias, and a spatially localized Detail Network to capture high-frequency stochastic convection, resolving transient singularities and preserving sharp boundaries. A subsequent Refiner module then dynamically reconstructs these decoupled multi-scale components into the final high-fidelity forecast. To address optimization instability, we introduce a multi-scale curriculum learning strategy that progressively shifts supervision from coarse scales to fine-grained details. Extensive experiments on the SEVIR and Shanghai Radar datasets demonstrate that WADEPre achieves state-of-the-art performance, yielding significant improvements in capturing extreme thresholds and maintaining structural fidelity. Our code is available at \url{https://github.com/sonderlau/WADEPre}.
\end{abstract}





\maketitle

\section{Introduction}

\begin{figure}[htbp]
    \centering
    \includegraphics[width=\columnwidth]{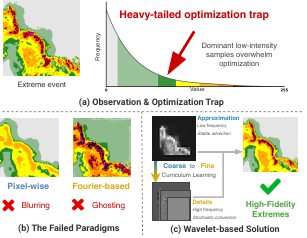}
    \caption{\textbf{Breaking the Forecasting Dilemma.} (a) The heavy-tailed distribution creates an optimization trap that ignores rare extremes. (b) Existing paradigms fail: pixel-wise methods blur details, while Fourier methods introduce ghosting artifacts. (c) WADEPre resolves this via wavelet decomposition, achieving sharp and accurate extremes.}
    \label{fig.bg}
\end{figure}

Severe convective storms, characterized by torrential precipitation, hail, and destructive winds, pose significant threats to public safety and economic stability~\cite{dgmr, nwp_nowcasting_chanllenges}. Accurate high-resolution precipitation nowcasting is crucial for disaster mitigation and urban hydrology management~\cite{schultzCanDeepLearning2021, Busker2025PrecipValue, Tafferner2012}. However, Numerical Weather Prediction (NWP) suffers from ``spin-up'' latency and high computational costs~\cite{nwp_nowcasting_chanllenges}, while optical flow methods (e.g., PySTEPS~\cite{pysteps}) often fail to capture nonlinear evolution. Therefore, reliably predicting extreme events remains a formidable challenge due to the heavy-tailed distribution. This creates a fundamental \textit{optimization dilemma}: standard pixel-wise objectives (e.g., Mean Squared Error) are statistically dominated by abundant low-intensity samples~\cite{zhaoLossFunc2017}. Consequently, models tend to regress towards the mean, systematically blurring out rare but high-impact extreme events~\cite{jennifer2025Heavytailed}. This fundamental \textit{optimization trap} caused by the heavy-tailed distribution is visualized in Figure~\ref{fig.bg} (Panel a).

To address the limitations of pixel-wise regression, the field has shifted toward spectral-domain modeling. Recent approaches have incorporated spectral losses (e.g., FFT-based objectives~\cite{facl}) to capture global patterns. However, a critical misalignment persists: these models still operate primarily in the pixel space, forcing the network to implicitly learn complex mappings from spatial inputs to spectral targets, often averaging out high-frequency textures during training~\cite{bonavitaLimitations2024}. Furthermore, while Fourier-domain models (e.g., AlphaPre~\cite{AlphaPre2025}) explicitly model frequencies, they are fundamentally constrained by the \textit{Heisenberg uncertainty principle}~\cite{mallat1999wavelet}. Fourier bases are globally localized in frequency but non-localized in space, making them inherently unsuitable for resolving transient, spatially localized convective cells, which are essential for extreme nowcasting. As illustrated in Figure~\ref{fig.bg} (Panel b), these existing paradigms fail to capture clear extremes: pixel-wise methods blur details, while Fourier transforms introduce ghosting artifacts.

Existing extreme precipitation nowcasting is thus constrained by three scientific challenges:

\begin{enumerate}
    \item \textbf{C1: Regression Bias in Heavy-tailed Distributions}. Low-intensity background noise dominates gradients from extreme events in standard optimization, leading to systematic attenuation and blurring of high-intensity cores~\cite{watsonMachineLearningApplications2022}.
    \item \textbf{C2: Spatial Localization Failure in the Spectral Domain}. Capturing the sharp boundaries of convective cells requires a representation that is both time- and frequency-localized~\cite{cnn_failed_spectral}. Fourier transforms fail to preserve this spatial locality, causing ghosting artifacts and structural incoherence.
    \item \textbf{C3: Spectral-Physical Inconsistency and Optimization Instability}. Modeling frequency bands independently creates a ``spectral barrier'', often resulting in reconstructed fields that lack physical consistency~\cite{facl}. Furthermore, directly learning high-frequency fluctuations without guidance is unstable, leading to optimization divergence~\cite{zhaoLossFunc2017}.
\end{enumerate}

To address these intrinsic limitations, we propose \textbf{WADEPre} (\textbf{WA}velet-based \textbf{D}ecomposition Model for \textbf{E}xtreme \textbf{Pre}cipitation Nowcasting), as shown in Figure~\ref{fig.bg} (Panel c). This physics-aware framework shifts the modeling process to the wavelet domain. To address the three challenges above (C1-C3), our solution incorporates three specific designs:
(1) \textbf{Approximation Network (A-Net) for Deterministic Advection (Solving C1):} Instead of a single regression, we use the Discrete Wavelet Transform (DWT) to isolate the stable low-frequency component. We propose an \textit{Approximation Network} to explicitly model deterministic motion trends, preventing heavy-tailed extremes from skewing the background optimization and resolving the regression-to-the-mean bias.
(2) \textbf{Details Network (D-Net) for Stochastic Convection (Solving C2):} To address the non-locality of Fourier methods, we design a \textit{Details Network} that operates on high-frequency wavelet coefficients. By leveraging wavelets' multi-scale spatial locality, D-Net explicitly captures transient convective cells and sharp gradients, thereby preserving structural fidelity and resolving ghosting artifacts.
(3) \textbf{Physics-aware Refiner with Curriculum Learning (Solving C3):} To bridge the spectral barrier and stabilize training, we propose a \textit{Refiner} module coupled with a \textit{Multi-Scale Curriculum Learning} strategy. The Refiner harmonizes the decoupled branches to enforce physical consistency, while the curriculum strategy gradually shifts supervision from coarse scales to fine-grained textures, ensuring robust convergence.

The main contributions of this work are summarized below:

\begin{enumerate}
    \item \textbf{Identification of Dual-Domain Bottlenecks:} We identify the \textit{regression bias} (leading to blurring) arising from heavy-tailed distributions and the \textit{spatial localization failures} (causing ghosting) in Fourier methods as the dual bottlenecks restricting extreme nowcasting.
    \item \textbf{Wavelet-based Disentanglement Model:} We propose \textbf{WADEPre}, which explicitly decomposes precipitation into stable deterministic advection and transient stochastic convection. These decoupled streams are harmonized by a Refiner to enforce physical consistency, thereby preventing high-frequency extremes from being oversmoothed by background trends.
    \item \textbf{Stable Multi-Scale Curriculum Optimization:} We introduce a \textit{coarse-to-fine} curriculum learning strategy. By leveraging wavelets' hierarchical properties, we progressively inject high-frequency supervision, thereby resolving the optimization instability inherent in extreme value prediction.
    \item \textbf{SOTA Performance in Extreme Event Forecasting:} Extensive experiments on the SEVIR and Shanghai Radar datasets demonstrate that WADEPre establishes a new state-of-the-art, specifically producing substantial improvements in Critical Success Index (CSI) at high thresholds while preserving excellent structural fidelity (SSIM).
\end{enumerate}

\section{Related Works}

\subsection{Data-Driven Meteorological Forecasting}

Deep learning has transformed weather forecasting, shifting from numerical integration to data-driven pattern recognition~\cite{pangu, graphcast, fuxi}. In precipitation nowcasting, early work framed the task as spatiotemporal sequence prediction. Recurrent Neural Networks (RNNs) pioneered this field, and ConvLSTM~\cite{ConvLSTM2015} and PredRNN~\cite{PredRNN2017, PredRNN++2018} introduced convolutional structures to capture spatiotemporal correlations. Recently, to mitigate the computational overhead of recurrence, simplified encoder-decoder architectures such as SimVP~\cite{gao2022simvp}, TAU~\cite{TAU2023}, and MAU~\cite{MAU} have emerged, achieving competitive performance while improving inference efficiency.

Despite these architectural advancements, these deterministic models inherently optimize pixel-wise objective functions such as Mean Squared Error (MSE). From a statistical perspective, minimizing MSE assumes a unimodal Gaussian target distribution, which ill suits the chaotic, multi-modal nature of precipitation~\cite{watsonMachineLearningApplications2022,video_beyond_mse}. Consequently, to reduce error, these models tend to regress toward a blurred average of possible future states, leading to severe suppression of high-frequency extreme events~\cite{cnn_failed_spectral, Yang2023}.

\subsection{Frequency-domain Forecasting}

To address the receptive field limitations of local convolutions, researchers have increasingly adopted frequency-domain modeling. The Fourier Neural Operator (FNO)~\cite{fno} pioneered this approach by solving PDEs (Partial Differential Equations) in the spectral domain. Building on this, AlphaPre~\cite{AlphaPre2025} used FFT-based disentanglement to model global evolution. Recently, WaveC2R~\cite{wavec2r} demonstrated the efficacy of wavelets in robust satellite-radar fusion, signaling a growing consensus on the utility of multi-resolution spectral modeling for meteorological boundaries.

However, Fourier methods are constrained by their global basis functions; a single coefficient perturbation affects the entire domain~\cite{wavelet_analysis}, making them suboptimal for localized singularities like sharp convective boundaries. This leads to Gibbs phenomena~\cite{gibbs} and spatial leakage, a trade-off dictated by the \textit{Heisenberg uncertainty principle}~\cite{mallat1999wavelet} which precludes simultaneous high resolution in both spatial and frequency domains.

\subsection{Physics-Informed and Generative Learning}

Bridging deep learning and atmospheric physics has become a critical research frontier. Hybrid frameworks, such as PhyDNet~\cite{PhyDNet2020} and EarthFarseer~\cite{Earthfarseer2024}, embed PDEs into recurrent cells to model physical dynamics. Recent studies~\cite{das2024hybrid} further explore coupling numerical solvers with neural networks to enhance robustness. In parallel, to address regression-to-the-mean blurriness, generative modeling has gained prominence. DGMR~\cite{dgmr} used GANs (Generative Adversarial Networks) to sharpen predictions, while subsequent diffusion-based models like NowcastNet~\cite{NowcastNet2023}, PreDiff~\cite{PreDiff2023}, and DiffCast~\cite{DiffCast2024} have achieved superior texture synthesis by iteratively refining Gaussian noise.

Practical deployment faces an accuracy-latency trade-off~\cite{PreDiff2023}. Physics-informed methods with soft regularization may produce inconsistent states~\cite{NowcastNet2023}. Generative models pose challenges: diffusion techniques require expensive sampling~\cite{DiffCast2024}, while GANs suffer from instability and mode collapse~\cite{dgmr}, limiting reliability.

\section{Methodology}

\subsection{Preliminaries: Wavelet Decomposition}

The 2D Discrete Wavelet Transform (DWT) recursively decomposes an input signal $\boldsymbol{A}^{l-1}$ (initially $\boldsymbol{A}^0 = \boldsymbol{X}$) into multi-resolution frequency sub-bands using low-pass ($L$) and high-pass ($H$) filters~\cite{mallat1999wavelet}. It applies separable 1D convolutions along the spatial dimensions, followed by dyadic downsampling ($\downarrow 2$), to yield a coarse approximation $\boldsymbol{A}^l$ and three detail components $\{\boldsymbol{D}_h^l, \boldsymbol{D}_v^l, \boldsymbol{D}_d^l\}$ that capture horizontal, vertical, and diagonal textures at level $l$:

\begin{equation}
\begin{cases}
\begin{aligned}
\boldsymbol{A}^{l} &= (L \otimes L) \boldsymbol{A}^{l-1} \downarrow 2 \\
\boldsymbol{D}^{l}_h &= (L \otimes H) \boldsymbol{A}^{l-1} \downarrow 2 \\
\boldsymbol{D}^{l}_v &= (H \otimes L) \boldsymbol{A}^{l-1} \downarrow 2 \\
\boldsymbol{D}^{l}_d &= (H \otimes H) \boldsymbol{A}^{l-1} \downarrow 2
\end{aligned}
\end{cases}
\end{equation}
where $\otimes$ denotes separable convolution. This hierarchical representation explicitly decomposes large-scale storm skeletons (captured in $\boldsymbol{A}^l$) from fine-grained intensity fluctuations (isolated in the detail coefficients).

\subsection{Problem Definition}

Following previous works~\cite{DiffCast2024, Earthformer2022, AlphaPre2025}, we formulate precipitation nowcasting as a sequence-to-sequence prediction problem. Given a sequence of radar observations $\boldsymbol{X} = \{ \boldsymbol{X}_{t-T_\text{in}+1}, \dots, \boldsymbol{X}_{t}\} \in \mathbb{R}^{T_\text{in} \times C \times H \times W}$, where $T_\text{in}$ denotes the input horizon, $C$ represents the number of channels (typically $C=1$ for radar echo), and $H, W$ denote the spatial resolution, the objective is to learn a mapping function $f_\theta$ with learnable parameters $\theta$ to predict the future sequence $\boldsymbol{Y} = \{\boldsymbol{Y}_{t+1}, \dots, \boldsymbol{Y}_{t+T_\text{out}}\} \in \mathbb{R}^{T_\text{out} \times C \times H \times W}$ for the subsequent $T_\text{out}$ steps. This process is formally expressed as $\boldsymbol{Y} = f_\theta(\boldsymbol{X})$.

\subsection{The Proposed WADEPre Model}

We formalize the problem of extreme precipitation nowcasting as a spatiotemporal sequence forecasting task in the wavelet domain. Let $\boldsymbol{X}_\text{seq}$ denote the input radar sequence. We apply a DWT to decompose it into a low-frequency approximation component ($\boldsymbol{A}_\text{seq}$) and a set of hierarchical high-frequency detail components ($\boldsymbol{D}_\text{seq}$).

\begin{figure*}[htbp]
  \centering
  \includegraphics[width=\linewidth]{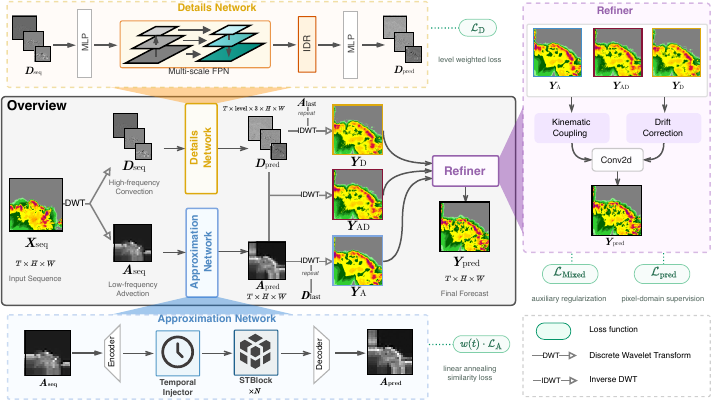}
  \caption{\textbf{Schematic overview of the WADEPre architecture}. The input sequence is decomposed via DWT into approximation ($\boldsymbol{A}_\text{seq}$) and details ($\boldsymbol{D}_\text{seq}$) coefficients. These components are processed by the dedicated \textit{Approximation Network} (Encoder-Mixer-Decoder) and \textit{Details Network} (Multi-scale FPN), respectively. The predicted coefficients are reconstructed via IDWT and fused by the \textit{Refiner} to generate the final forecast $\boldsymbol{Y}_\text{pred}$. The green capsules indicate the loss functions applied during training.}
\label{fig.wadepre_architecture}
\end{figure*}

As shown in Figure~\ref{fig.wadepre_architecture}, WADEPre processes these components through three specialized modules. The \textbf{Approximation Network} is designed to capture deterministic global trends, ensuring the stability of large-scale advection. In parallel, the \textbf{Details Network} focuses on modeling stochastic local fluctuations to preserve high-frequency extremes often lost in standard regression. Finally, the \textbf{Refiner} integrates these multi-scale outputs to enforce spectral consistency and correct spatial alignment artifacts.

\subsection{Approximation Network}
\label{sec.a_net}

\begin{figure}[htbp]
  \centering
  \includegraphics[width=\columnwidth]{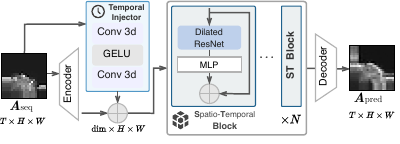}
  \caption{\textbf{Architecture of the Approximation Network (A-Net).} Designed to model deterministic low-frequency advection. The network employs a \textit{Temporal Injector} (via 3D Convolution) to extract inter-frame dynamics from the input sequence $\boldsymbol{A}_\text{seq}$. The core evolution is driven by stacked \textit{Spatio-Temporal Blocks} (STBlocks), which capture synoptic-scale spatial dependencies without loss of resolution.}
  \label{fig.a_net}
\end{figure}

The Approximation Network (A-Net) captures the slow, deterministic evolution of the precipitation field, representing physical cloud systems, as illustrated in Figure~\ref{fig.a_net}. Given input wavelet approximation $\boldsymbol{A}_\text{seq}$, A-Net predicts future state $\boldsymbol{A}_\text{pred}$.

To effectively model long-term dependencies while preserving spatial consistency, we propose a \textit{Spatio-Temporal Dilated Injection} architecture. The workflow comprises three stages: (1) Encoding and Temporal Injection, (2) Spatio-Temporal Dilated Evolution, and (3) Decoding and Stationary-Texture Reconstruction.

\textbf{(1) Encoding and Temporal Injection}. The low-frequency coefficients exhibit high spatial redundancy and strong temporal continuity. We first map the sequence into a high-dimensional latent space using a 2D Convolutional Encoder. We treat the time dimension T as channels to compress the temporal dimension: 

\begin{equation}
\boldsymbol{Z}_\text{enc}  = \text{Conv2d}(\boldsymbol{A}_\text{seq})  \in \mathbb{R}^{\text{dim} \times H \times W}.
\end{equation}

Simultaneously, to explicitly preserve the inter-frame dynamic properties (e.g., translation and rotation of air masses), we introduce a \textit{Temporal Injector}. This module employs 3D Convolutions to extract volumetric spatiotemporal features from the sequence:

\begin{equation}
\boldsymbol{Z}_\text{inj} = \text{Conv3d}(\boldsymbol{A}_\text{seq}) \in \mathbb{R}^{ \text{dim} \times H \times W}.
\end{equation}

The injected features ($\boldsymbol{Z}_\text{inj}$) are then added to the encoded features ($\boldsymbol{Z}_\text{enc} $) and fused via a $1 \times 1$ Convolution to form the initial hidden state $\boldsymbol{Z}_0$.

\textbf{(2) Spatio-Temporal Dilated Evolution.} The core evolution is driven by a stack of \textit{Spatio-Temporal Blocks} (STBlocks). To enable the network to perceive large-scale advection patterns without losing resolution, we decompose the mixing process:

\begin{itemize} 

\item \textit{Spatial Dilated Mixing}: We employ a Dilated ResNet~\cite{dilated_resnet} structure. By exponentially increasing the dilation rate $d_k = 2^k$ at the $k$-th block ($k \in [0, N]$), the receptive field expands exponentially. This allows the network to capture global spatial dependencies while keeping the feature map size: 
\begin{equation}
\boldsymbol{Z}_k = \text{DilatedResNet}(\boldsymbol{Z}_{k-1}, d_k).
\end{equation}
\item \textit{Temporal Channel Mixing}: Implemented via $1 \times 1$ convolutions, this module operates exclusively along the channel dimension $\text{dim}$. Since temporal dynamics are encoded within these channels, the MLP facilitates dense interaction among latent time steps to propagate evolution information:
\begin{equation}
    \boldsymbol{Z}_{k+1} = \text{MLP}(\boldsymbol{Z}_{k}) + \boldsymbol{Z}_k .
\end{equation}
This operation models the transition dynamics while preserving the spatial structure.
\end{itemize}

\textbf{(3) Decoding and Stationary-Texture Reconstruction}. After $N$ blocks of evolution, a \textit{Decoder} projects the hidden features ($\boldsymbol{Z}_N$) back to the prediction horizon $T$, yielding the predicted approximation coefficients $\boldsymbol{A}_\text{pred} \in \mathbb{R}^{T \times H \times W}$.

To reconstruct the image-space background flow $\boldsymbol{Y}_A$, we adopt a \textit{Stationary Texture Assumption}. Since the high-frequency details ($\boldsymbol{D}$ coefficients) represent textures that move with the flow but change slowly in statistical distribution, we utilize the last observed detail frame from the input sequence ($\boldsymbol{D}_\text{seq}$) and repeat it across the prediction horizon: 

\begin{equation} 
\boldsymbol{Y}_A = \text{IDWT} \left(\boldsymbol{A}_\text{pred}, \; \text{Repeat}(\boldsymbol{D}_\text{last}, T) \right).
\end{equation}

This strategy ensures that the reconstructed background retains realistic textural sharpness while strictly following the predicted advection path defined by $\boldsymbol{A}_\text{pred}$.

\subsection{Details Network}
\label{sec.d_net}

While the A-Net captures the deterministic background flow, the Details Network (D-Net) models the volatile, high-frequency components of the precipitation field, specifically the rapid formation and dissipation of convective cells. These components correspond to the wavelet detail coefficients $\boldsymbol{D}_\text{seq} = \{ \boldsymbol{D}^l \}_{l=1}^{\text{level}}$ across multiple decomposition levels.

\begin{figure}[htbp]
\centering
\includegraphics[width=\columnwidth]{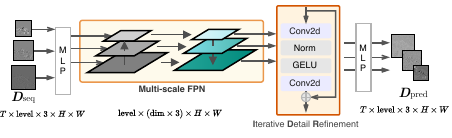}
\caption{\textbf{Architecture of the Detail Network (D-Net)}. Designed to resolve stochastic high-frequency convection. The network projects the detail coefficients $\boldsymbol{D}_\text{seq}$ into latent feature spaces via Temporal MLPs. A \textit{Feature Pyramid Network (FPN)} backbone facilitates bidirectional cross-scale energy transfer, while the proposed \textit{Iterative Detail Refinement (IDR)} module rectifies spectral inconsistencies to preserve sharp, localized boundaries in the final prediction.}
\label{fig.d_net}
\end{figure}

To resolve the challenge of predicting stochastic textures without spectral loss, we propose a \textit{Hierarchical Stochastic Refinement} architecture, shown in Figure~\ref{fig.d_net}. The network proceeds in four stages: (1) Temporal Projection, (2) Cross-Scale Spatial Interaction, (3) Iterative Detail Refinement, and (4) Decoding and Reconstruction.

\textbf{(1) Temporal Projection}. Unlike the A-Net, where spatial context dominates, the detail coefficients at each level ($\boldsymbol{D}_{\text{seq}}^{l}$) are highly sensitive to short-term fluctuations. A shared temporal MLP projects the input sequence into a latent feature space $\boldsymbol{V}$:

\begin{equation}
\boldsymbol{V} = \text{MLP} (\boldsymbol{D}^{l}_\text{seq}) \in \mathbb{R}^{\text{level} \times  ( \text{dim} \times 3) \times H_l \times W_l }.
\end{equation}

This projection compresses the temporal evolution into channel descriptors, preparing the features for spatial interaction.

\textbf{(2) Cross-Scale Spatial Interaction}. We employ a Feature Pyramid Network (FPN)~\cite{FPN2017} backbone to facilitate information exchange across frequency levels. Unlike standard FPNs, which generate a pyramid from a single input resolution, we adopt a \textit{parallel multi-scale input} strategy, in which the projected detail features at each level are directly injected into the corresponding FPN stage.

Once injected, the FPN orchestrates a bidirectional information exchange to harmonize features across scales:
\begin{itemize} 
	\item \textit{Bottom-up Aggregation:} High-frequency texture information from finer scales is propagated upward to enrich the semantic representation of coarser levels. 
	\item \textit{Top-down Guidance:} Coarse-scale structural context flows downward to guide the consistent evolution of fine-grained details.
\end{itemize}

This design ensures that the output features at each level serve as precursors for $\boldsymbol{D}_\text{pred}^l$. Integrate both the specific frequency characteristics of their native level and the contextual constraints from neighboring scales.

\textbf{(3) Iterative Detail Refinement (IDR)}. To further enhance the sharpness of extreme events, we introduce the \textit{Iterative Detail Refinement} (IDR) module. As shown in Figure~\ref{fig.d_net}, the IDR acts as a residual correction block applied at each pyramid level. It consists of a \textit{Conv2d-Norm-GELU-Conv2d} sequence designed to rectify spectral inconsistencies:

\begin{equation}
\boldsymbol{V}^{l}_\text{IDR} = \boldsymbol{V}^{l}_\text{FPN} + \text{IDR}(\boldsymbol{V}^{l}_\text{FPN}).
\end{equation}
where $\boldsymbol{V}^l_\text{FPN}$ is the ouput of FPN at level $l$. By adding the refined residuals back to the FPN output, the network explicitly learns to recover the high-frequency boundaries lost during downsampling.

\textbf{(4) Decoding and Reconstruction}. Finally, the refined features ($\boldsymbol{V}_\text{IDR}$) are mapped back to the temporal domain via a decoding MLP, yielding the predicted detail coefficients $\boldsymbol{D}_\text{pred}$. To reconstruct the high-frequency convective field $\boldsymbol{Y}_D$, we employ the IDWT. Similar to the strategy employed in the A-Net, we use the last observed approximation frame, $\boldsymbol{A}_\text{last}$, from $\boldsymbol{A}_\text{seq}$ as a static structural anchor. $\boldsymbol{A}_\text{last}$ is repeated across the horizon to align with the temporal dimension of $\boldsymbol{D}_\text{pred}$:

\begin{equation}
    \boldsymbol{Y}_D = \text{IDWT} \left( \text{Repeat}(\boldsymbol{A}_\text{last}, \; T), \boldsymbol{D}_\text{pred} \right).
\end{equation}

This design leverages $\boldsymbol{A}_\text{last}$ to provide the basic spatial layout, ensuring that the reconstructed $\boldsymbol{Y}_D$ explicitly captures the sharp, stochastic variations produced by the D-Net.

\subsection{Refiner}

While the A-Net and D-Net effectively decompose the evolution of distinct frequency bands, simply reconstructing the final forecast via IDWT is insufficient. As demonstrated in our ablation studies (Section~\ref{sec.ablation}), direct reconstruction often yields predictions lacking in physical consistency due to the spectral barrier between the two independent sub-networks. To address this, we propose the Refiner, a physics-aware harmonization module that rectifies spectral inconsistencies and aligns the forecast with natural atmospheric evolution.

Let $\boldsymbol{Y}_A$ and $\boldsymbol{Y}_D$ denote the partial reconstructions representing the low-frequency background energy and high-frequency convective energy, respectively. $\boldsymbol{Y}_{AD} = \text{IDWT} (\boldsymbol{A}_\text{pred}, \boldsymbol{D}_\text{pred}) $ denote the preliminary full reconstruction. The Refiner operates through two parallel physics-guided streams:

\begin{itemize}
    \item \textit{Kinematic Coupling ($f_{KC}$):} This module fuses the partial reconstructions ($\boldsymbol{Y}_A$ and $\boldsymbol{Y}_D$) via an initial $3\times3$ convolution followed by stacked residual blocks. By learning joint spatial representations, the network implicitly promotes kinematic coherence, spatially anchoring high-frequency convective cores to the low-frequency advection skeleton to resolve spectral inconsistencies.
    \item \textit{Drift Correction ($f_{DC}$):} To maintain temporal continuity, we explicitly calculate the deviation from the latest observation: $\boldsymbol{Y}_{AD} - \boldsymbol{X}_\text{last}$. Residual blocks process these differential features with a global skip connection. This residual correction mechanism mitigates accumulated trajectory errors, effectively serving as a soft inertial constraint to prevent implausible state jumps.
\end{itemize}

Finally, the feature maps from both streams are fused via a 2D convolutional layer with residual connections to the preliminary reconstruction $\boldsymbol{Y}_{AD}$. The final prediction $\boldsymbol{Y}_\text{pred}$ is formulated as:

\begin{equation}
\boldsymbol{Y}_\text{pred} = \boldsymbol{Y}_{AD} + \text{Conv2D}\left( [ f_{KC}(\boldsymbol{Y}_{A}, \boldsymbol{Y}_{D}), f_{DC}(\boldsymbol{Y}_{AD} - \boldsymbol{X}_\text{last}) ] \right).
\end{equation}
where $f_{KC}$ and $f_{DC}$ denote the Kinematic Coupling and Drift Correction, respectively, and $[ \cdot ]$ is the concatenation operation. 

This design ensures that the final output ($\boldsymbol{Y}_{AD}$) maintains the sharpness of extreme events (via the D-Net), the structural stability of the storm system (via the A-Net), and physical consistency (via the Drift Correction and Kinematic Coupling).

\subsection{Multi-Scale Curriculum Learning Strategy}

To effectively train the decomposition architecture while respecting the distinct physical roles of each component, we introduce a multi-scale curriculum learning strategy. This strategy uses dynamically weighted loss functions to guide the optimization process from coarse-scale motion to fine-scale intensity refinement. The specific application points of these losses are marked in Figure~\ref{fig.wadepre_architecture} with green capsules. The total objective function $\mathcal{L}_\text{total}$ is formulated as:

\begin{equation}
	\mathcal{L}_\text{total} = \mathcal{L}_{\text{pred}} +  w(t) \cdot \mathcal{L}_A + \lambda_D \cdot \mathcal{L}_D + \lambda_\text{Mixed} \cdot \mathcal{L}_{\text{Mixed}}.
\end{equation}
where $t$ denotes the current training steps. $\mathcal{L}_{\text{pred}} = \text{MSE}(\boldsymbol{Y}_{\text{pred}}, \boldsymbol{Y}_{\text{target}})$ directly supervises the final precipitation forecast in the pixel domain, and $\lambda_{D, \text{Mixed}}$ are the balancing factors.

 The approximation branch captures the large-scale advection trends (the storm's skeleton). To enforce structural alignment while allowing for local intensity variations, we employ the Zero-Normalized Cross-Correlation (ZNCC) loss:

\begin{equation}
\label{eq.loss_a}
	\mathcal{L}_A = 1 - \text{ZNCC} \left( \boldsymbol{A}_{\text{pred}}, \boldsymbol{A}_{\text{target}} \right).
\end{equation}

Here, the ZNCC metric measures structural similarity that is invariant to linear intensity transformations. To facilitate convergence, we apply a linear annealing schedule $w(t)$ that prioritizes coarse structure learning in the early stages:

\begin{equation}
\label{eq.loss_weight_a}
	w(t) = \max\left(1 - t /  T_{\text{decay}}, \; \lambda_{\text{min}} \right).
\end{equation}
where $T_{\text{decay}}$ determines the duration of the curriculum phase, and $\lambda_{\text{min}}$ is the minimum value.

The D-Net models intensity fluctuations across multiple scales. We define a level-weighted loss to balance the contribution of different frequency bands:

\begin{equation}
	\mathcal{L}_D = \sum_{l=1}^{\text{level}} \frac{1}{2^l} \cdot \text{MSE}(\boldsymbol{D}_{\text{pred}}^l, \boldsymbol{D}_{\text{target}}^l ).
\end{equation}
where $l$ denotes the decomposition level, this weight schedule prevents the model from overfitting to noise while ensuring the recovery of sharp intensity peaks.

To prevent the intermediate representations from diverging and to ensure collaborative learning across branches, we introduce an auxiliary regularization term:

\begin{equation}
\label{eq.loss_mixer}
\mathcal{L}_{\text{Mixed}} = \text{MSE}\left(  \left(\boldsymbol{Y}_{\text{A}} + \boldsymbol{Y}_{\text{D}} + \boldsymbol{Y}_{\text{AD}} \right) /  3, \; \boldsymbol{Y}_{\text{target}}\right).
\end{equation}

This term serves as a consistency constraint, encouraging the approximation branch ($\boldsymbol{Y}_A$), detail branch ($\boldsymbol{Y}_D$), and directly reconstructed branch ($\boldsymbol{Y}_{AD}$) to converge to the ground truth collectively.

\section{Experiments}

\subsection{Experimental Setup}

\subsubsection{Baselines}

To comprehensively evaluate our proposed approach, we select five representative baseline models for comparison: (1) ConvLSTM~\cite{ConvLSTM2015}, (2) MAU~\cite{MAU}, (3) SimVP~\cite{gao2022simvp}, (4) EarthFarseer~\cite{Earthfarseer2024}, and (5) AlphaPre~\cite{AlphaPre2025}. Following prior research~\cite{AlphaPre2025}, these baselines are categorized into two classifications: decomposition (type D) and non-decomposition (type ND).

\subsubsection{Dataset}
We conducted experiments on two precipitation datasets: SEVIR~\cite{SEVIR2020} covers the United States at 384 km $\times$ 384 km with observations at 5-minute intervals. A 10-minute resolution was employed. Shanghai Radar~\cite{shanghai_radar} records precipitation over an area of 501 km $\times$ 501 km, with data inputs of 6-minute intervals and outputs at 12-minute intervals. The parameters $T_\text{in}$ and $T_\text{out}$ were set to 6, and the input data were resized to $128 \times 128$ for both datasets. All models were trained from scratch. Further details are provided in Appendix~\ref{appdix.dataset}.

\subsubsection{Metrics}

 We use the Root Mean Squared Error (RMSE) for numerical prediction error, and the Structural Similarity Index Measure (SSIM)~\cite{ssim} for structural preservation. To evaluate forecasting skill, we compute the Heidke Skill Score (HSS), which measures the difference from a random prediction. The Critical Success Index (CSI) is evaluated at a specific threshold. We calculate the mean of the six thresholds (CSI-M) and select two thresholds (CSI-H and CSI-E) for validating the extreme value benchmark. See Appendix~\ref{appdix.metrics} for more details.

\subsubsection{Implementation Details} 

Baselines utilize official repositories configured with default settings. The comprehensive training procedures and hyperparameters are detailed in Appendix~\ref{appdix.implementation}.

 \subsection{Performance Comparison}

Table~\ref{tab.performance_mean} summarizes the quantitative performance averaged across the six forecast lead times. WADEPre establishes a new state-of-the-art on both the SEVIR and Shanghai Radar benchmarks, consistently outperforming baselines across critical metrics, including CSI-M, extreme event indicators (e.g., CSI-219, CSI-40), HSS, and SSIM.

\begin{table*}[htbp]
  \caption{Quantitative comparison averaged across all six lead times on the SEVIR and Shanghai Radar dataset. Type D and ND denote the decomposition and non-decomposition models, respectively. $\uparrow$ indicates higher is better, $\downarrow$ indicates lower is better. The best results are highlighted in \textbf{bold}, and the second-best are \underline{underlined}.}
  \label{tab.performance_mean}
  \centering
  \resizebox{\textwidth}{!}{
    \begin{tabular}{l r cccccc cccccc}
    \toprule
    \multirow{2}{*}{Model} & \multirow{2}{*}{Type} & \multicolumn{6}{c}{SEVIR} & \multicolumn{6}{c}{Shanghai Radar} \\
    \cmidrule(lr){3-8} \cmidrule(lr){9-14}
    & & CSI-M $\uparrow$ & CSI-181 $\uparrow$ & CSI-219 $\uparrow$ & RMSE $\downarrow$ & HSS $\uparrow$ & SSIM $\uparrow$ 
    & CSI-M $\uparrow$ & CSI-35 $\uparrow$ & CSI-40 $\uparrow$ & RMSE $\downarrow$ & HSS $\uparrow$ & SSIM $\uparrow$ \\
    \midrule

    ConvLSTM & ND    
    & 0.355974 & 0.155084 & 0.041291 & 1.290777 & 0.445585 & 0.717261
    & 0.253558 & 0.052567 & 0.001231 & 3.033739 & 0.337086 & \underline{0.770083} \\

    MAU & ND       
    & 0.378454 & 0.179911 & 0.078185 & 1.290873 & 0.477095 & 0.719202 
    & 0.346315 & 0.249759 & 0.126814 & 3.234246 & 0.473638 & 0.736891 \\
    
    SimVP & ND       
    & 0.391180 & 0.203362 & 0.073078 & 1.244711 & 0.496391 & 0.668615 
    & 0.322941 & 0.191222 & 0.074395 & 3.165812 & 0.413999 & 0.738400 \\
    
    EarthFarseer & D
    & 0.394133 & 0.203624 & 0.064953 & 1.238125 & 0.494665 & 0.545065 
    & 0.362593 & 0.258890 & 0.051279 & \underline{2.607779} & 0.477972 & 0.498071 \\
    
    AlphaPre & D    
    & \underline{0.408885} & \underline{0.224541} & \underline{0.082268} & \textbf{1.207027} & \underline{0.512415} & \underline{0.749047} 
    & \underline{0.409432} & \underline{0.303714} & \underline{0.191909} & 2.663889 & \underline{0.542150} & 0.726093 \\

    WADEPre & D
    & \textbf{0.416419} & \textbf{0.238489} & \textbf{0.115865} & \underline{1.232280} & \textbf{0.526560} & \textbf{0.754846} 
    & \textbf{0.421976} & \textbf{0.317689} & \textbf{0.201965} & \textbf{2.595196} & \textbf{0.550064} & \textbf{0.770512} \\
    
    \bottomrule
    \end{tabular}
  }
\end{table*}

\begin{figure}[htbp]
  \centering
  \includegraphics[width=\columnwidth]{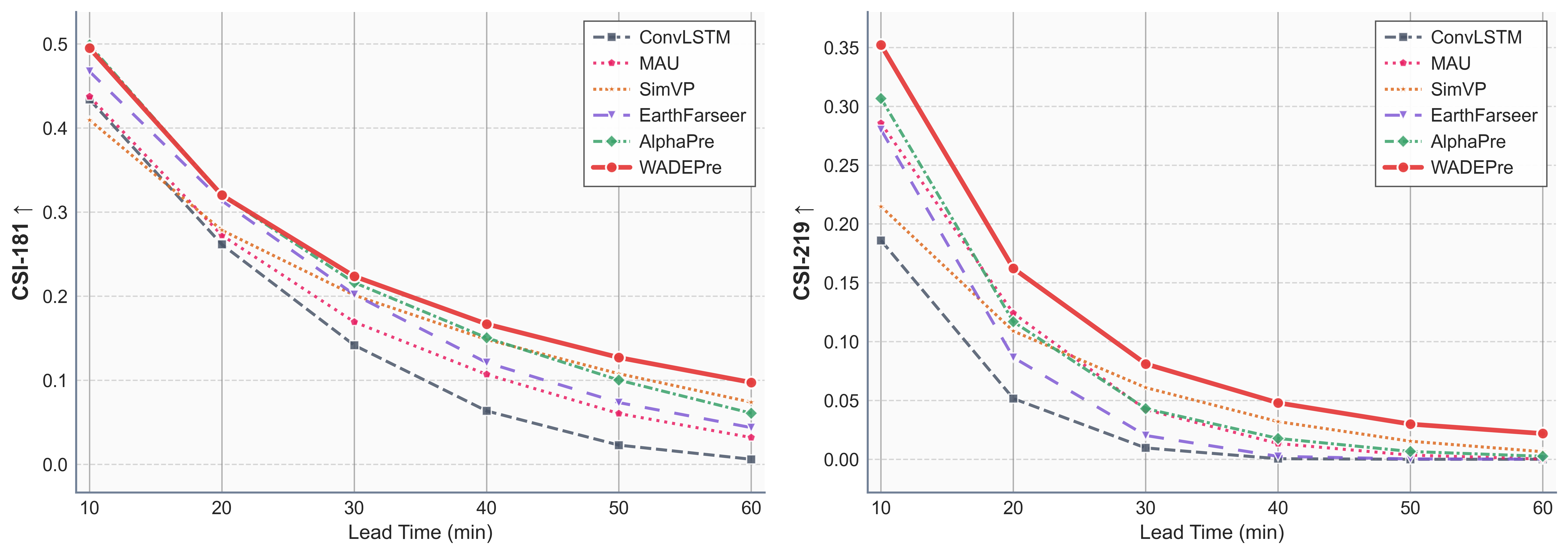}
  \caption{\textbf{Temporal evolution of forecast skill for extreme events on the SEVIR.} The curves visualize frame-wise CSI scores at high thresholds (CSI-181 and CSI-219) from 10 to 60 minutes. WADEPre demonstrates better long-term robustness than baselines.}
  \label{fig.csi_comparison}
\end{figure}

To further analyze the model's robustness over longer lead times, Figure~\ref{fig.csi_comparison} illustrates the frame-wise performance evolution for heavy (CSI-181) and extreme (CSI-219) precipitation events on the SEVIR dataset. We summarize the comparison of performance as follows:

\begin{itemize}
    \item \textit{Forecasting Skill}: WADEPre achieves SOTA performance across aggregated CSI and HSS metrics. Notably, our significant lead at high thresholds (as detailed in Figure~\ref{fig.csi_comparison}) demonstrates the efficacy of the proposed architecture.

    \item \textit{Structural Fidelity}: Achieving the highest SSIM scores on both datasets demonstrates WADEPre's topological superiority. This corroborates that the Wavelet-based Decomposition method, coupling the coarse-grained reconstruction of the A-Net with the multi-scale texture refinement of the D-Net, accurately preserves the spatial distribution and structural coherence of future precipitation fields.
   
    \item \textit{Numerical Error}: WADEPre's marginally higher RMSE on SEVIR reflects the \textit{double penalty} effect~\cite{doublepenalty2025}, where sharp predictions are penalized for minor spatial misalignments. Unlike pixel-wise baselines that blur outputs or AlphaPre that lacks wavelet-like spatial localization, WADEPre preserves singularities. Given the substantial gains in CSI and HSS, this trade-off is essential for meteorological value.
\end{itemize}

\subsection{Visualization and Case Study}

\begin{figure}[ht]
  \centering
  \includegraphics[width=\columnwidth]{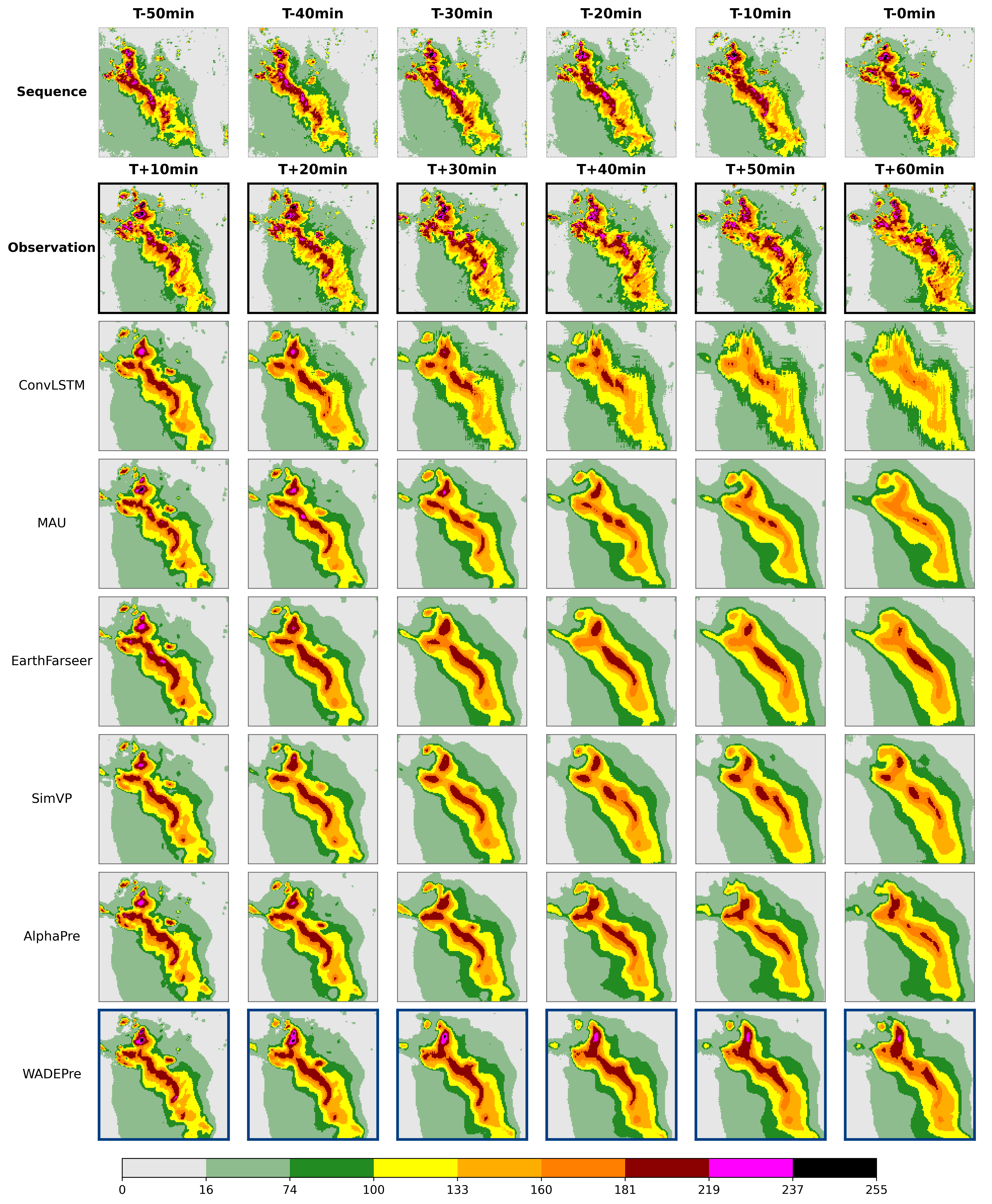}
  \caption{\textbf{Qualitative visualization of a high-intensity flash flood event}. Comparison of forecasts from T+10 to T+60 min. While baselines suffer from severe spectral smoothing and rapid intensity attenuation, causing the convective core to dissipate into background noise, WADEPre (blue frames) exhibits superior morphological consistency. It successfully preserves the sharp boundaries and high-intensity peaks (pink/black regions).}
  \label{fig.case_study}
\end{figure}

To evaluate the model's ability to capture severe weather events, we visualize forecast results for a linear squall line, which is characterized by an elongated, high-intensity echo band with sharp convective boundaries, in Figure~\ref{fig.case_study}.

Conventional pixel-wise regression models (ConvLSTM, MAU, EarthFarseer, and SimVP) exhibit significant diffusive behavior. As the prediction lead time extends beyond T+30 min, the sharp gradients of the squall line are smoothed, causing the system to lose its linear organization and degrade the high-intensity core into a diffuse cloud. While AlphaPre preserves partial intensity, it suffers from structural fragmentation, failing to maintain the continuity of the convective line.

In contrast, WADEPre demonstrates superior kinematic consistency. It successfully decomposes large-scale advection from local intensity changes, thereby preserving the morphological integrity of the squall line and accurately predicting the propagation of the high-intensity leading edge up to T+60 minutes.

\subsection{Ablation Study}
\label{sec.ablation}

In this section, we investigate the individual contributions of the model's core architectural components and the proposed multi-scale curriculum learning strategy.

\subsubsection{Contribution of Architectural Components}

We first evaluate the three core modules: the Approximation Network (A-Net), the Detail Network (D-Net), and the Refiner. The ablation variants are defined as follows: (1) w/o A-Net: The A-Net is removed, and the Refiner takes $\boldsymbol{A}_\text{seq}$ directly as the approximation prior; (2) w/o D-Net: The D-Net is removed, and $\boldsymbol{D}_\text{seq}$ serves as the substitute for the detail prior; (3) w/o Refiner: The final prediction is generated directly via the IDWT of the predicted coefficients.

 \begin{table}[htbp]
 \caption{Ablation study results for the components of WADEPre on the SEVIR dataset. $\uparrow$ indicates higher is better, $\downarrow$ indicates lower is better. The best results are highlighted in \textbf{bold}, and the second-best are \underline{underlined}.}
  \label{tab.ablation_components}
 \centering
 
 \resizebox{\columnwidth}{!}{
 \begin{tabular}{l c  c c c c c}
 \hline
  Variant  & CSI-M $\uparrow$ & CSI-181 $\uparrow$ & CSI-219 $\uparrow$ & RMSE $\downarrow$ & HSS $\uparrow$ & SSIM $\uparrow$ \\
 \hline
 	w/o A-Net & 0.190881 & 0.019060 & 0.003977 & 1.794504 & 0.244787 & 0.594154 \\
 	w/o D-Net & \underline{0.365966} & \underline{0.176532} & \underline{0.080723} & 1.402748 & \underline{0.471082} & 0.672986 \\
 	w/o Refiner & 0.362865 & 0.154449 & 0.063399 & \underline{1.383330} & 0.464147 & \underline{0.677992} \\
   WADEPre & \textbf{0.416419} & \textbf{0.238489} & \textbf{0.115865} & \textbf{1.232280} & \textbf{0.526560} & \textbf{0.754846}  \\
 \hline
 \end{tabular}
 }
 \end{table}
 
Table~\ref{tab.ablation_components} confirms A-Net as the backbone; removing it causes a ~54\% CSI-M collapse, confirming global advection as a prerequisite. Omitting D-Net significantly degrades performance on the CSI-181 and CSI-219 metrics, highlighting its role in resolving singularities. The Refiner is crucial for spectral components; without it, direct superposition causes spectral leakage and ringing, reducing reconstruction quality.

\subsubsection{Impact of Loss Functions and Curriculum Learning}

We further analyze the impact of our multi-scale curriculum learning strategy. The variants include: (1) w/o $w(t)$: removes the dynamic weighting schedule; (2) w/o $\mathcal{L}_\text{A}$ and (3) w/o $\mathcal{L}_\text{D}$: remove the intermediate supervision for the A-Net and D-Net branches, respectively; (4) $\mathcal{L}_\text{Mixed}$: excludes the consistency regularization term.

Table~\ref{tab.ablation_loss} confirms the effectiveness of the curriculum strategy. Removing dynamic weighting $w(t)$ causes the largest performance drop, indicating that prioritizing low-frequency foundations avoids bottlenecks. Omitting intermediate supervision ($\mathcal{L}_\text{A}$ and $\mathcal{L}_\text{D}$) degrades performance by removing targeted guidance. In contrast, $\mathcal{L}_\text{Mixed}$ is crucial for regularizing spectral-pixel consistency.
 
   \begin{table}[htbp]
  \caption{Ablation study results for the loss function of WADEPre on the SEVIR dataset. $\uparrow$ indicates higher is better, $\downarrow$ indicates lower is better. The best results are highlighted in \textbf{bold}, and the second-best are \underline{underlined}.}
  \label{tab.ablation_loss}
 \centering
 \resizebox{\columnwidth}{!}{
 \begin{tabular}{l c  c c c c c}
 \hline
  Variant  & CSI-M $\uparrow$ & CSI-181 $\uparrow$ & CSI-219 $\uparrow$ & RMSE $\downarrow$ & HSS $\uparrow$ & SSIM $\uparrow$ \\
 \hline
 	w/o $w(t)$ & 0.383020 & 0.188383 & 0.092796 & 1.287580 & 0.483875 & 0.704201  \\
 	w/o $\mathcal{L}_\text{A}$ & 0.395002 & 0.203690 & 0.087669 & 1.270574 & 0.499125 & \underline{0.752970}  \\
 	w/o $\mathcal{L}_\text{D}$ & 0.387864 & 0.191810 & 0.086628 & 1.277584& 0.488795 & 0.750401 \\
 	w/o $\mathcal{L}_\text{Mixed}$ & \underline{0.400809} & \underline{0.218417} & \underline{0.100016} & \underline{1.265551} & \underline{0.508016} & 0.735994 \\
   WADEPre & \textbf{0.416419} & \textbf{0.238489} & \textbf{0.115865} & \textbf{1.232280} & \textbf{0.526560} & \textbf{0.754846} \\
 \hline
 \end{tabular}
 }
 \end{table}

\section{Conclusions}

We address the persistent regression-to-the-mean dilemma in extreme precipitation nowcasting with WADEPre, a physics-aware wavelet-based decomposition model. By decomposing spatiotemporal evolution into deterministic advection (\textit{A-Net}) and stochastic fluctuations (\textit{D-Net}), and mixing them with a \textit{Refiner}, our model mitigates spectral bias and ensures structural coherence. Ablation studies demonstrate the necessity of this hierarchical design and emphasize the contribution of each module. Evaluations on the SEVIR and Shanghai Radar benchmarks show that WADEPre sets a new state-of-the-art, especially in predicting hazardous extremes. Currently, the model is in trial operation and assessment in Zhejiang Province, demonstrating its practical utility for real-world forecasting. By combining data-driven learning with spectral signal processing, WADEPre offers a promising approach to enhancing the physical reliability of AI-based weather forecasting models.

\section*{Limitations and Ethical Considerations}

\noindent \textbf{Limitations and Future Work.} WADEPre currently lacks strict thermodynamic consistency and incurs high computational costs due to multi-scale transforms. Future iterations will address these limitations by incorporating multivariate meteorological priors to enforce physical constraints and by integrating diffusion models to quantify probabilistic uncertainty.

\noindent \textbf{Ethical Considerations.} We use publicly available datasets~\cite{SEVIR2020,shanghai_radar} that contain no personally identifiable information (PII) and involve no human subjects, thereby exempting this research from Institutional Review Board (IRB) oversight. Concerning fairness, geographic bias remains a concern; deploying this technology in underrepresented regions requires local fine-tuning to ensure equitable outcomes. Lastly, to prevent misuse arising from overreliance, this system is intended solely as a decision-support tool to complement professionals' expertise, rather than as a substitute for established operational protocols.

\begin{acks}
 This work was supported by the National Natural Science Foundation of China (U2342218), the ``Pioneer'' and ``Leading Goose'' R\&D Program of Zhejiang (Grant No. 2024C03256), the Joint Funds of the Zhejiang Provincial Natural Science Foundation of China (Grant No. LZJMY24D050007), and the China Meteorological Administration (Grant No. FPZJ2025-053).
\end{acks}

\bibliographystyle{ACM-Reference-Format}
\bibliography{reference}

@article{bonavitaLimitations2024,
author = {Bonavita, Massimo},
title = {On Some Limitations of Current Machine Learning Weather Prediction Models},
journal = {Geophysical Research Letters},
volume = {51},
number = {12},
pages = {e2023GL107377},
doi = {10.1029/2023GL107377},
year = {2024}
}

@inproceedings{gao2022simvp,
author    = {Gao, Zhangyang and Tan, Cheng and Wu, Lirong and Li, Stan Z.},
title     = {{SimVP}: Simpler Yet Better Video Prediction},
booktitle = {Proceedings of the {IEEE/CVF} Conference on Computer Vision and Pattern Recognition ({CVPR})},
month     = {June},
year      = {2022},
pages     = {3170-3180},
url = {https://openaccess.thecvf.com/content/CVPR2022/papers/Gao_SimVP_Simpler_Yet_Better_Video_Prediction_CVPR_2022_paper.pdf},
}

@inproceedings{ConvLSTM2015,
author = {Shi, Xingjian and Chen, Zhourong and Wang, Hao and Yeung, Dit-Yan and Wong, Wai-kin and Woo, Wang-chun},
title = {Convolutional LSTM Network: a machine learning approach for precipitation nowcasting},
year = {2015},
address = {Cambridge, MA, USA},
booktitle = {Proceedings of the 29th International Conference on Neural Information Processing Systems},
pages = {802-–810},
url = {https://dl.acm.org/doi/10.5555/2969239.2969329},
}

@inproceedings{Earthfarseer2024,
author = {Wu, Hao and Liang, Yuxuan and Xiong, Wei and Zhou, Zhengyang and Huang, Wei and Wang, Shilong and Wang, Kun},
title = {{Earthfarseer}: versatile spatio-temporal dynamical systems modeling in one model},
booktitle = {Proceedings of the AAAI Conference on Artificial Intelligence},
pages={15906-15914},
volume={38},
  year={2024},
  doi = {10.1609/aaai.v38i14.29521},
}

@inproceedings{MAU,
 author = {Chang, Zheng and Zhang, Xinfeng and Wang, Shanshe and Ma, Siwei and Ye, Yan and Xinguang, Xiang and Gao, Wen},
 booktitle = {Proceedings of the International Conference on Neural Information Processing Systems ({NeurIPS})},
 address = {Red Hook, NY, USA},
 title = {MAU: a motion-aware unit for video prediction and beyond},
 url = {https://dl.acm.org/doi/10.5555/3540261.3542325},
 articleno = {2064},
 numpages = {13},
 year = {2021}
}

@inproceedings{SEVIR2020,
 author = {Veillette, Mark and Samsi, Siddharth and Mattioli, Chris},
 booktitle = {Advances in {Neural Information Processing Systems}},
 editor = {H. Larochelle and M. Ranzato and R. Hadsell and M.F. Balcan and H. Lin},
 pages = {22009--22019},
 publisher = {Curran Associates, Inc.},
 title = {{SEVIR} : A Storm Event Imagery Dataset for Deep Learning Applications in Radar and Satellite Meteorology},
 url = {https://proceedings.neurips.cc/paper_files/paper/2020/file/fa78a16157fed00d7a80515818432169-Paper.pdf},
 volume = {33},
 year = {2020}
}

@article{shanghai_radar,
author = {Chen, Lei and Cao, Yuan and Ma, Leiming and Zhang, Junping},
title = {A Deep Learning-Based Methodology for Precipitation Nowcasting With Radar},
journal = {Earth and Space Science},
volume = {7},
number = {2},
pages = {e2019EA000812},
doi = {10.1029/2019EA000812},
url = {https://agupubs.onlinelibrary.wiley.com/doi/abs/10.1029/2019EA000812},
year = {2020}
}

@inproceedings{FPN2017,
  author={Lin, Tsung-Yi and Dollár, Piotr and Girshick, Ross and He, Kaiming and Hariharan, Bharath and Belongie, Serge},
  booktitle={2017 IEEE Conference on Computer Vision and Pattern Recognition (CVPR)},
  title={Feature Pyramid Networks for Object Detection},
  pages={936-944},
  year={2017},
  doi={10.1109/CVPR.2017.106}
}

@book{mallat1999wavelet,
  title={A Wavelet Tour of Signal Processing},
  author={Mallat, Stephane},
  isbn={9780124666061},
  lccn={99065087},
  series={Electronics \& Electrical},
  url={https://books.google.com.hk/books?id=yW2kut44AsMC},
  year={1999},
  publisher={Elsevier Science}
}

@inproceedings{TAU2023,
author={Tan, Cheng and Gao, Zhangyang and Wu, Lirong and Xu, Yongjie and Xia, Jun and Li, Siyuan and Li, Stan Z.},
booktitle={2023 IEEE/CVF Conference on Computer Vision and Pattern Recognition (CVPR)},
title={{Temporal Attention Unit}: Towards Efficient Spatiotemporal Predictive Learning},
year={2023},
pages={18770-18782},
doi={10.1109/CVPR52729.2023.01800}
}

@inproceedings{PreDiff2023,
  title={{PreDiff}: Precipitation Nowcasting with Latent Diffusion Models},
  author={Gao, Zhihan and Shi, Xingjian and Han, Boran and Wang, Hao and Jin, Xiaoyong and Maddix, Danielle C and Zhu, Yi and Li, Mu and Wang, Bernie},
  booktitle= {Proceedings of the 37th International Conference on Neural Information Processing Systems},
  address = {Red Hook, NY, USA},
  pages = {78621 -- 7865},
  year={2023},
  url = {https://dl.acm.org/doi/10.5555/3666122.3669561},
}

@inproceedings{PredRNN2017,
author = {Wang, Yunbo and Long, Mingsheng and Wang, Jianmin and Gao, Zhifeng and Yu, Philip S.},
title = {PredRNN: recurrent neural networks for predictive learning using spatiotemporal LSTMs},
year = {2017},
address = {Red Hook, NY, USA},
pages = {879–888},
numpages = {10},
url = {https://dl.acm.org/doi/10.5555/3294771.3294855}
}

@inproceedings{Earthformer2022,
  title={Earthformer: exploring space-time transformers for earth system forecasting},
  author = {Gao, Zhihan and Shi, Xingjian and Wang, Hao and Zhu, Yi and Wang, Yuyang and Li, Mu and Yeung, Dit-Yan},
  booktitle = {Proceedings of the 36th International Conference on Neural Information Processing Systems},
  pages={25390--25403},
  year={2022},
  url = {https://dl.acm.org/doi/10.5555/3600270.3602111},
  address = {Red Hook, NY, USA},
}

@inproceedings{PredRNN++2018,
  title = 	 {{P}red{RNN}++: Towards A Resolution of the Deep-in-Time Dilemma in Spatiotemporal Predictive Learning},
  author =       {Wang, Yunbo and Gao, Zhifeng and Long, Mingsheng and Wang, Jianmin and Yu, Philip S},
  booktitle = 	 {Proceedings of the 35th International Conference on Machine Learning},
  pages = 	 {5123--5132},
  year = 	 {2018},
  volume = 	 {80},
  month = 	 {Jul},
  publisher =  {{PMLR}},
  url = 	 {https://proceedings.mlr.press/v80/wang18b.html},
}

@article{dgmr,
author = {Ravuri, Suman and Lenc, Karel and Willson, Matthew and Kangin, Dmitry and Lam, Remi and Mirowski, Piotr and Fitzsimons, Megan and Athanassiadou, Maria and Kashem, Sheleem and Madge, Sam and Prudden, Rachel and Mandhane, Amol and Clark, Aidan and Brock, Andrew and Simonyan, Karen and Hadsell, Raia and Robinson, Niall and Clancy, Ellen and Arribas, Alberto and Mohamed, Shakir},
	date = {2021/09/01},
	doi = {10.1038/s41586-021-03854-z},
	id = {Ravuri2021},
	isbn = {1476-4687},
	journal = {Nature},
	number = {7878},
	pages = {672--677},
	title = {Skilful precipitation nowcasting using deep generative models of radar},
	volume = {597},
	year = {2021},
}

@inproceedings{PhyDNet2020,
  booktitle={2020 IEEE/CVF Conference on Computer Vision and Pattern Recognition (CVPR)},
  title={Disentangling Physical Dynamics From Unknown Factors for Unsupervised Video Prediction},
  author={Guen, Vincent Le and Thome, Nicolas},
  pages={11471-11481},
  year={2020},
  doi={10.1109/CVPR42600.2020.01149}
}

@article{NowcastNet2023,
	author = {Zhang, Yuchen and Long, Mingsheng and Chen, Kaiyuan and Xing, Lanxiang and Jin, Ronghua and Jordan, Michael I. and Wang, Jianmin},
	doi = {10.1038/s41586-023-06184-4},
	isbn = {1476-4687},
	journal = {Nature},
	number = {7970},
	pages = {526--532},
	title = {Skilful nowcasting of extreme precipitation with {NowcastNet}},
	volume = {619},
	year = {2023},
}

@INPROCEEDINGS{DiffCast2024,
  author={Yu, Demin and Li, Xutao and Ye, Yunming and Zhang, Baoquan and Luo, Chuyao and Dai, Kuai and Wang, Rui and Chen, Xunlai},
  booktitle={2024 {IEEE/CVF} Conference on Computer Vision and Pattern Recognition ({CVPR})},
  title={DiffCast: A Unified Framework via Residual Diffusion for Precipitation Nowcasting},
  year={2024},
  pages={27758--27767},
  doi={10.1109/CVPR52733.2024.02622}
}

@InProceedings{AlphaPre2025,
      author={Lin, Kenghong and Zhang, Baoquan and Yu, Demin and Feng, Wenzhi and Chen, Shidong and Gao, Feifan and Li, Xutao and Ye, Yunming},
      booktitle={2025 IEEE/CVF Conference on Computer Vision and Pattern Recognition (CVPR)},
      title={AlphaPre: Amplitude-Phase Disentanglement Model for Precipitation Nowcasting},
      year      = {2025},
      doi={10.1109/CVPR52734.2025.01662},
      pages={17841-17850},
  }

@article{Yang2023,
  author = {Yang, Shangshang and Yuan, Huiling},
  title = {A Customized Multi-Scale Deep Learning Framework for Storm Nowcasting},
  journal = {Geophysical Research Letters},
  volume = {50},
  number = {13},
  pages = {e2023GL103979},
  doi = {10.1029/2023GL103979},
  year = {2023}
  }

@article{das2024hybrid,
	author = {Das, Puja and Posch, August and Barber, Nathan and Hicks, Michael and Duffy, Kate and Vandal, Thomas and Singh, Debjani and Werkhoven, Katie van and Ganguly, Auroop R.},
	journal = {npj Climate and Atmospheric Science},
	number = {1},
	year = {2024},
	pages = {},
	publisher = {Springer Science and Business Media LLC},
	title = {Hybrid physics-{AI} outperforms numerical weather prediction for extreme precipitation nowcasting},
	volume = {7},
	doi = {10.1038/s41612-024-00834-8}
  }

@ARTICLE{zhaoLossFunc2017,
  author={Zhao, Hang and Gallo, Orazio and Frosio, Iuri and Kautz, Jan},
  journal={{IEEE} Transactions on Computational Imaging},
  title={{Loss Functions for Image Restoration With Neural Networks}},
  year={2017},
  volume={3},
  number={1},
  pages={47--57},
  doi={10.1109/TCI.2016.2644865}
}

@article{pangu,
  title    = {Accurate Medium-Range Global Weather Forecasting with 3D Neural Networks},
  author   = {Bi, Kaifeng and Xie, Lingxi and Zhang, Hengheng and Chen, Xin and Gu, Xiaotao and Tian, Qi},
  year     = {2023},
  month    = jul,
  journal  = {Nature},
  volume   = {619},
  number   = {7970},
  pages    = {533--538},
  issn     = {0028-0836, 1476-4687},
  doi      = {10.1038/s41586-023-06185-3},
  url      = {https://www.nature.com/articles/s41586-023-06185-3},
  urldate  = {2023-09-27},
  langid   = {english},
  lccn     = {1},
}

@article{fuxi,
  author  = {Chen, Lei and Zhong, Xiaohui and Zhang, Feng and others},
  title   = {{FuXi}: A cascade machine learning forecasting system for 15-day global weather forecast},
  journal = {npj Clim. Atmos. Sci.},
  year    = {2023},
  volume  = {6},
  number  = {1},
  pages   = {190},
  doi     = {10.1038/s41612-023-00512-1}
}

@article{graphcast,
  author = {Remi Lam  and Alvaro Sanchez-Gonzalez  and Matthew Willson  and Peter Wirnsberger  and Meire Fortunato  and Ferran Alet  and Suman Ravuri  and Timo Ewalds  and Zach Eaton-Rosen  and Weihua Hu  and Alexander Merose  and Stephan Hoyer  and George Holland  and Oriol Vinyals  and Jacklynn Stott  and Alexander Pritzel  and Shakir Mohamed  and Peter Battaglia },
  title = {Learning skillful medium-range global weather forecasting},
  journal = {Science},
  year    = {2023},
  volume  = {382},
  number  = {6677},
  pages   = {1416--1421},
  doi     = {10.1126/science.adi2336}
}

@article{schultzCanDeepLearning2021,
  title     = {Can Deep Learning Beat Numerical Weather Prediction?},
  author    = {Schultz, M. G. and Betancourt, C. and Gong, B. and Kleinert, F. and Langguth, M. and Leufen, L. H. and Mozaffari, A. and Stadtler, S.},
  year      = {2021},
  month     = {02},
  journal = {Philosophical Transactions of the Royal Society A: Mathematical, Physical and Engineering Sciences},
  volume = {379},
  number = {2194},
  pages = {20200097},
  publisher = {Royal Society},
  doi       = {10.1098/rsta.2020.0097},
  url       = {https://royalsocietypublishing.org/doi/full/10.1098/rsta.2020.0097},
  urldate   = {2024-01-16}
}

@inbook{Tafferner2012,
  author    = {Tafferner, Arnold and Forster, Caroline},
  title     = {Weather Nowcasting and Short Term Forecasting},
  booktitle = {Atmospheric Physics: Background -- Methods -- Trends},
  year      = {2012},
  publisher = {Springer Berlin Heidelberg},
  address   = {Berlin, Germany},
  pages     = {363--380},
  doi       = {10.1007/978-3-642-30183-4\_22},
  isbn={978-3-642-30183-4},
}

@inproceedings{video_beyond_mse,
      title={Deep multi-scale video prediction beyond mean square error},
      author={Michael, Mathieu and Camille, Couprie and Yann, LeCun},
      year={2016},
      address = {San Juan, Puerto Rico},
      booktitle = {the International Conference on Learning Representations ({ICLR})},
      doi = {10.48550/arXiv.1511.05440},
}

@inproceedings{cnn_failed_spectral,
  author={Durall, Ricard and Keuper, Margret and Keuper, Janis},
  booktitle={IEEE/CVF Conference on Computer Vision and Pattern Recognition (CVPR)},
  title={Watch Your Up-Convolution: CNN Based Generative Deep Neural Networks Are Failing to Reproduce Spectral Distributions},
  year={2020},
  address = {Seattle, WA, USA},
  pages={7887-7896},
  doi={10.1109/CVPR42600.2020.00791}
}

@article{ssim,
  author={Zhou Wang and Bovik, A.C. and Sheikh, H.R. and Simoncelli, E.P.},
  journal={IEEE Transactions on Image Processing},
  title={Image quality assessment: from error visibility to structural similarity},
  year={2004},
  volume={13},
  number={4},
  pages={600--612},
  doi={10.1109/TIP.2003.819861}
}

@inproceedings{facl,
author = {Yan, Chiu-Wai and Foo, Shi Quan and Trinh, Van Hoan and Yeung, Dit-Yan and Wong, Ka-Hing and Wong, Wai-Kin},
title = {Fourier amplitude and correlation loss: beyond using L2 loss for skillful precipitation nowcasting},
year = {2024},
booktitle = {38th International Conference on Neural Information Processing Systems},
numpages = {35},
address = {Red Hook, NY, USA},
url = {https://dl.acm.org/doi/10.5555/3737916.3741089},
}

@article{nwp_nowcasting_chanllenges,
author = "Juanzhen Sun and Ming Xue and James W. Wilson and Isztar Zawadzki and Sue P. Ballard and Jeanette Onvlee-Hooimeyer and Paul Joe and Dale M. Barker and Ping-Wah Li and Brian Golding and Mei Xu and James Pinto",
title = "Use of NWP for Nowcasting Convective Precipitation: Recent Progress and Challenges",
journal = "Bulletin of the American Meteorological Society",
year = "2014",
publisher = "American Meteorological Society",
volume = "95",
number = "3",
doi = "10.1175/BAMS-D-11-00263.1",
pages=      "409 -- 426",
}

@article{watsonMachineLearningApplications2022,
  title = {Machine Learning Applications for Weather and Climate Need Greater Focus on Extremes},
  author = {Watson, Peter A G},
  year = {2022},
  journal = {Environmental Research Letters},
  volume = {17},
  number = {11},
  pages = {111004},
  publisher = {IOP Publishing},
  doi = {10.1088/1748-9326/ac9d4e},
}

@article{wavelet_analysis,
  author={Daubechies, I.},
  journal={IEEE Transactions on Information Theory},
  title={The wavelet transform, time-frequency localization and signal analysis},
  year={1990},
  volume={36},
  number={5},
  pages={961-1005},
  doi={10.1109/18.57199}
}

@article{gibbs,
author = {Gottlieb, David and Shu, Chi-Wang},
title = {On the Gibbs Phenomenon and Its Resolution},
journal = {SIAM Review},
volume = {39},
number = {4},
pages = {644--668},
year = {1997},
doi = {10.1137/S0036144596301390},
}

@inproceedings{wavec2r,
      title={WaveC2R: Wavelet-Driven Coarse-to-Refined Hierarchical Learning for Radar Retrieval},
      author={Chunlei Shi and Han Xu and Yinghao Li and Yi-Lin Wei and Yongchao Feng and Yecheng Zhang and Dan Niu},
      year={2025},
      booktitle={Proceedings of the Conference Association for the Advancement of Artificial Intelligence ({AAAI})},
      address = {Singapore},
      doi = {10.48550/arXiv.2511.17558}
}

@article{pysteps,
  author  = {Pulkkinen, S. and Nerini, D. and P\'erez Hortal, A. A. and Velasco-Forero, C. and Seed, A. and Germann, U. and Foresti, L.},
  title   = {Pysteps: an open-source Python library  for probabilistic precipitation nowcasting (v1.0)},
  journal = {Geosci. Model Dev.},
  volume  = {12},
  year    = {2019},
  number  = {10},
  pages   = {4185--4219},
  url     = {https://gmd.copernicus.org/articles/12/4185/2019/},
  doi     = {10.5194/gmd-12-4185-2019}
}

@inproceedings{doublepenalty2025,
  title     = {Fixing the Double Penalty in Data-Driven Weather Forecasting Through a Modified Spherical Harmonic Loss Function},
  author    = {Subich, Christopher and Husain, Syed Zahid and Separovic, Leo and Yang, Jing},
  booktitle = {International Conference on Machine Learning ({ICML})},
  year      = {2025},
  address = {Vancouver, BC, Canada},
  url       = {https://icml.cc/virtual/2025/poster/44912}
}

@inproceedings{fno,
  title={Fourier Neural Operator for Parametric Partial Differential Equations},
  author={Li, Zongyi and Kovachki, Nikola and Azizzadenesheli, Kamyar and Liu, Burigede and Bhattacharya, Kaushik and Stuart, Andrew and Anandkumar, Anima},
  booktitle={International Conference on Learning Representations ({ICLR})},
  year={2021},
  url={https://openreview.net/forum?id=c8P9NQVtmnO}
}

@article {Busker2025PrecipValue,
      author = "Tim Busker and Bart van den Hurk and Hans de Moel and Jeroen C. J. H. Aerts",
      title = "The Value of Precipitation Forecasts to Anticipate Floods",
      journal = "Bulletin of the American Meteorological Society",
      year = "2025",
      publisher = "American Meteorological Society",
      address = "Boston MA, USA",
      volume = "106",
      number = "3",
      doi = "10.1175/BAMS-D-24-0073.1",
      pages= "E473--E491",
}

@Article{jennifer2025Heavytailed,
AUTHOR = {Wang, Hsing-Jui and Merz, Ralf and Basso, Stefano},
TITLE = {Constructing a geography of heavy-tailed flood distributions: insights from common streamflow dynamics},
JOURNAL = {Hydrology and Earth System Sciences},
VOLUME = {29},
YEAR = {2025},
NUMBER = {6},
PAGES = {1525--1548},
DOI = {10.5194/hess-29-1525-2025}
}

@inproceedings{dilated_resnet,
  title={Dilated Residual Networks},
  author={Yu, Fisher and Koltun, Vladlen and Funkhouser, Thomas},
  booktitle={Proceedings of the IEEE Conference on Computer Vision and Pattern Recognition (CVPR)},
  year={2017},
  pages={636--644},
  doi={10.1109/CVPR.2017.75}
}

\appendix

\section{Baselines}

\subsection{ConvLSTM}
ConvLSTM~\cite{ConvLSTM2015} is a pioneering work that extends the fully connected LSTM to the spatiotemporal domain by incorporating convolutional structures into recurrent transitions. We select it as the foundational benchmark to quantify the performance leap from classical pixel-wise recurrent architectures to modern spectral designs. 

\subsection{MAU}
MAU (Motion-Aware Unit)~\cite{MAU} introduces a dedicated attention mechanism to capture predictive motion dynamics between consecutive frames. We include MAU as a representative of the motion-centric recurrent model class. This comparison allows us to rigorously evaluate whether implicit motion modeling via attention mechanisms can rival our explicit Kinematic Coupling strategy.

\subsection{SimVP}
SimVP~\cite{gao2022simvp} simplifies spatiotemporal modeling into a pure encoder-decoder CNN framework, achieving competitive results on various benchmarks by effectively capturing long-range dependencies. We include SimVP as a representative baseline of modern CNN-based methods. The comparison demonstrates that, unlike SimVP, which may struggle with high-frequency textures, WADEPre successfully maintains the structural integrity of sharp convective boundaries.

\subsection{EarthFarseer}
EarthFarseer~\cite{Earthfarseer2024} is a recent Transformer-based model that integrates earth-specific physical constraints into the attention mechanism to model spatiotemporal turbulent flows. We select it as a representative of physics-informed Transformers to evaluate the effectiveness of soft physical constraints.

\subsection{AlphaPre}
AlphaPre~\cite{AlphaPre2025} is state-of-the-art in frequency-domain nowcasting, using FFT to separate amplitude and phase, disentangling intensity from motion. It serves as a baseline to compare the global Fourier Transform's utility with our local Wavelet Transform.

\section{Metrics}
\label{appdix.metrics}

\textbf{Critical Success Index (CSI)}. CSI (also known as the Thread Score) evaluates the forecasting skill for specific precipitation thresholds. It is defined as:
\begin{equation}
    \text{CSI} = \frac{TP}{TP + FP + FN}
\end{equation}
where TP, FP, and FN represent the number of true positives, false positives, and false negatives, respectively, after binarizing the input tensors.

We report CSI at three different intensity levels to evaluate performance across varying rainfall severities: mean intensity (CSI-M), high intensity (CSI-H), and extreme intensity (CSI-E). All six thresholds and high intensity (H) and extreme intensity (E) are shown in Table~\ref{tab.metric_thresholds}. A higher CSI value indicates better performance.

 \begin{table}[htbp]
  \caption{Thresholds on CSI and HSS.}
  \label{tab.metric_thresholds}
 \centering
 \resizebox{\columnwidth}{!}{
 \begin{tabular}{l c c c}
 \hline
  Dataset  & Thresholds & High Intensity (H) & Extreme Intensity (E)\\
 \hline
SEVIR & \{ 16, 74, 133, 160, 181, 219 \} & 181 & 219\\
Shanghai & \{ 20, 30, 35, 40 \} & 35 & 40\\
 \hline
 \end{tabular}
 }
 \end{table}

\textbf{Root Mean Squared Error (RMSE)}. RMSE measures pixel-level differences between predicted frames and ground truth. It is defined as:
\begin{equation}
    \text{RMSE} = \sqrt{\frac{1}{N} \sum_{i=1}^{N} (y_i - \hat{y}_i)^2}
\end{equation}
where $N$ is the total number of pixels, $y_i$ represents the ground truth value, and $\hat{y}_i$ represents the predicted value. Lower RMSE values indicate better predictive accuracy.

To evaluate RMSE in the physical domain, we denormalize the model outputs before calculation. For the SEVIR dataset, predictions are mapped from $[0, 1]$ back to pixel integers $[0, 255]$ and subsequently converted to physical quantities ($\text{kg} / \text{m}^2$) via Equation~\ref{eq:vil_transform}. Similarly, for the Shanghai Radar dataset, the outputs are denormalized to the original reflectivity range of $0 \sim 70$ dBZ.

\textbf{Heidke Skill Score (HSS)}. The HSS quantifies forecast accuracy relative to random chance, accounting for correct negatives which are dominant in sparse precipitation fields:
\begin{equation}
    \text{HSS} = \frac{2 \times (TP \times TN - FP \times FN)}{(TP + FN) (FN + TN) + (TP + FP)(FP + TN)}
\end{equation}
where TP, FN, FP, and TN represent the number of true positives, false negatives, false positives, and true negatives,  respectively, after binarizing the input tensors.

HSS ranges from 0 to 1, averaged across the six defined intensity thresholds (see Table~\ref{tab.metric_thresholds}), with a higher value indicating better forecasting skill.

\textbf{Structural Similarity Index Measure (SSIM)}. SSIM assesses the perceptual quality of the predicted images by evaluating luminance, contrast, and structural information. It is calculated as:
\begin{equation}
    \text{SSIM}(x, y) = \frac{(2\mu_x\mu_y + C_1)(2\sigma_{xy} + C_2)}{(\mu_x^2 + \mu_y^2 + C_1)(\sigma_x^2 + \sigma_y^2 + C_2)}
\end{equation}
where $x$ and $y$ are two images, $\mu$ and $\sigma$ represent the mean and variance of the images, and $\sigma_{xy}$ is the covariance. SSIM values range from 0 to 1, with higher values indicating greater structural similarity.

\section{Dataset}
\label{appdix.dataset}

On both datasets, $T_\text{in}$ and $T_\text{out}$ are set to 6, and $C$ is set to 1 (omitted in notations for brevity).

\subsection{SEVIR}

We use the Vertically Integrated Liquid (VIL) product from the SEVIR dataset~\cite{SEVIR2020}. Provided initially at $384 \times 384$ ($1$ km pixel pitch), the data are bilinearly downsampled to $128 \times 128$ and sampled at 10-minute intervals. The raw pixel values $x \in [0, 254]$ (with 255 denoting missing data) are normalized to the range $[0, 1]$ for training.

To ensure physical significance during evaluation, the RMSE metric is computed on the denormalized physical quantities $y$ (in units of $\text{kg}/\text{m}^2$), as specified by the official protocol:

\begin{equation}
y = \begin{cases} 
	0 & x \le 5 \\
   (x - 2) / 90.66 & 5 < x \le 18 \\
	\exp\left( (x - 83.9) / 38.9 \right) & 18 < x \le 254
	\end{cases}
\label{eq:vil_transform} 
\end{equation}

\subsection{Shanghai Radar}

We utilize the Shanghai Radar dataset~\cite{shanghai_radar}, collected by the Shanghai Central Meteorological Observatory (SCMO), which consists of Composite Reflectivity (CR) sequences from the Yangtze River Delta. The raw data spans a $460 \times 460$ grid covering a physical region of $460 \text{km} \times 398 \text{km}$, with reflectivity values ranging from 0 to 70 dBZ. 

In our experimental setting, the input sequence consists of observations sampled at 6-minute intervals, while the forecasting target consists of observations sampled at 12-minute intervals. Following the preprocessing protocol, the radar maps are resized to $128 \times 128$ and normalized to the $[0, 1]$ range. Consistent with SEVIR, the RMSE metric is computed on the denormalized values (restored to the 0--70 dBZ range) to reflect physical prediction errors accurately.

\section{Quantitative Analysis}

Complementing the focused analysis of extreme events (CSI-181 and CSI-219) presented in Figure~\ref{fig.csi_comparison}, this section offers a comprehensive evaluation of the model's performance across all metrics. The temporal progression of six principal indicators: CSI-Mean, CSI-High, CSI-Extreme, RMSE, HSS, and SSIM, across six lead times.

\begin{figure}[htbp]
  \centering
  \includegraphics[width=\columnwidth]{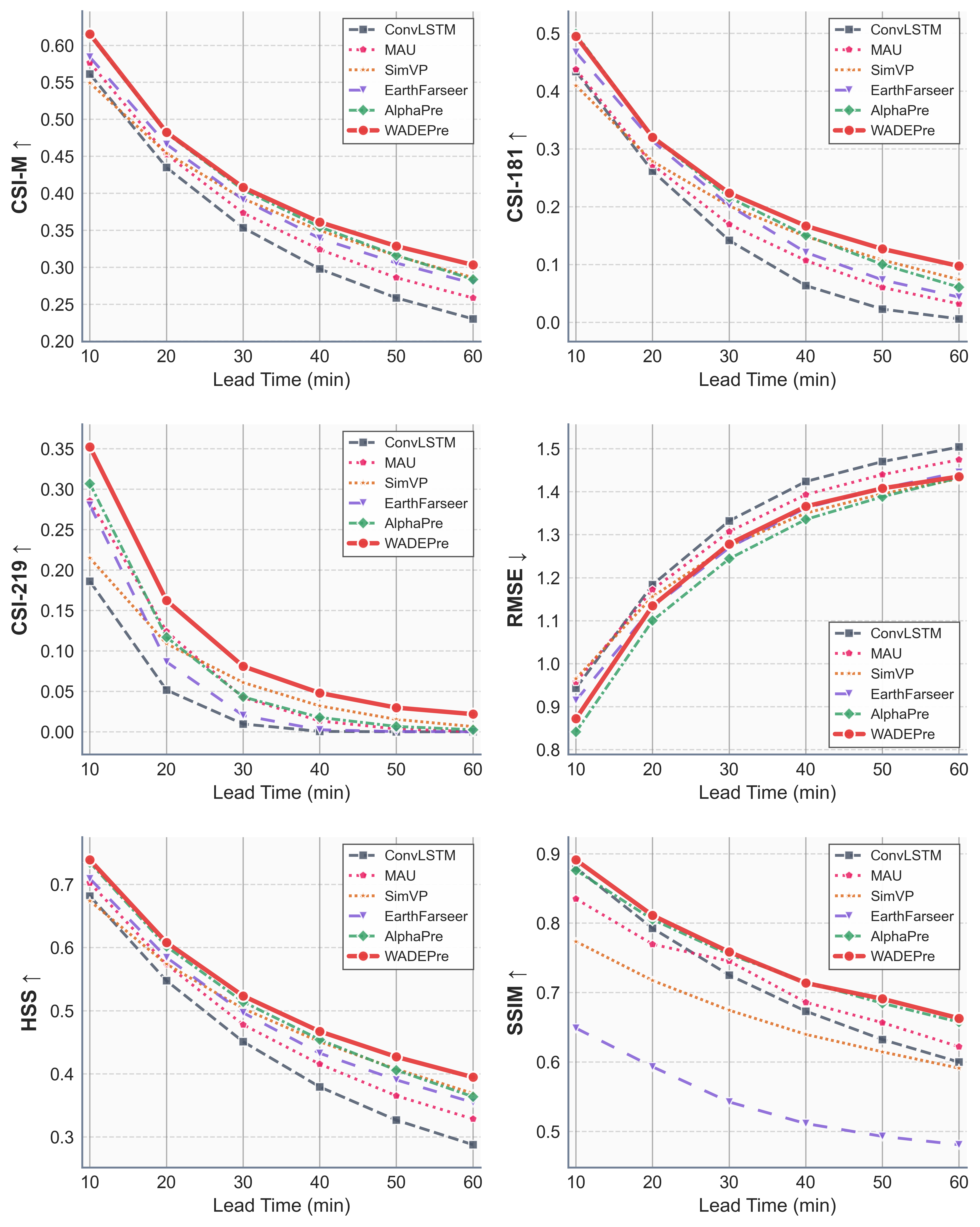}
  \caption{\textbf{Comprehensive temporal evaluation on the SEVIR benchmark}. The curves depict the frame-wise evolution of six metrics over the 60-minute prediction lead time. WADEPre (red) maintains the highest scores across all skill metrics, particularly in high-intensity regimes (CSI-181 and CSI-219) and in structural similarity (SSIM), compared to the second-best method AlphaPre.}
\label{fig.metrics_all_sevir}
\end{figure}

Figure~\ref{fig.metrics_all_sevir} illustrates the comparison of frame-wise performance on the SEVIR dataset. WADEPre consistently demonstrates superior results, especially in later stages of prediction. Conversely, baseline models such as ConvLSTM and SimVP exhibit rapid decline in performance as lead time extends, whereas WADEPre maintains a more stable trajectory. Notably, in terms of CSI-219 and SSIM, our model effectively reduces smoothing effects, preserving intricate storm details that are typically lost by pixel-wise regression baselines.

\begin{figure}[htbp]
  \centering
  \includegraphics[width=\columnwidth]{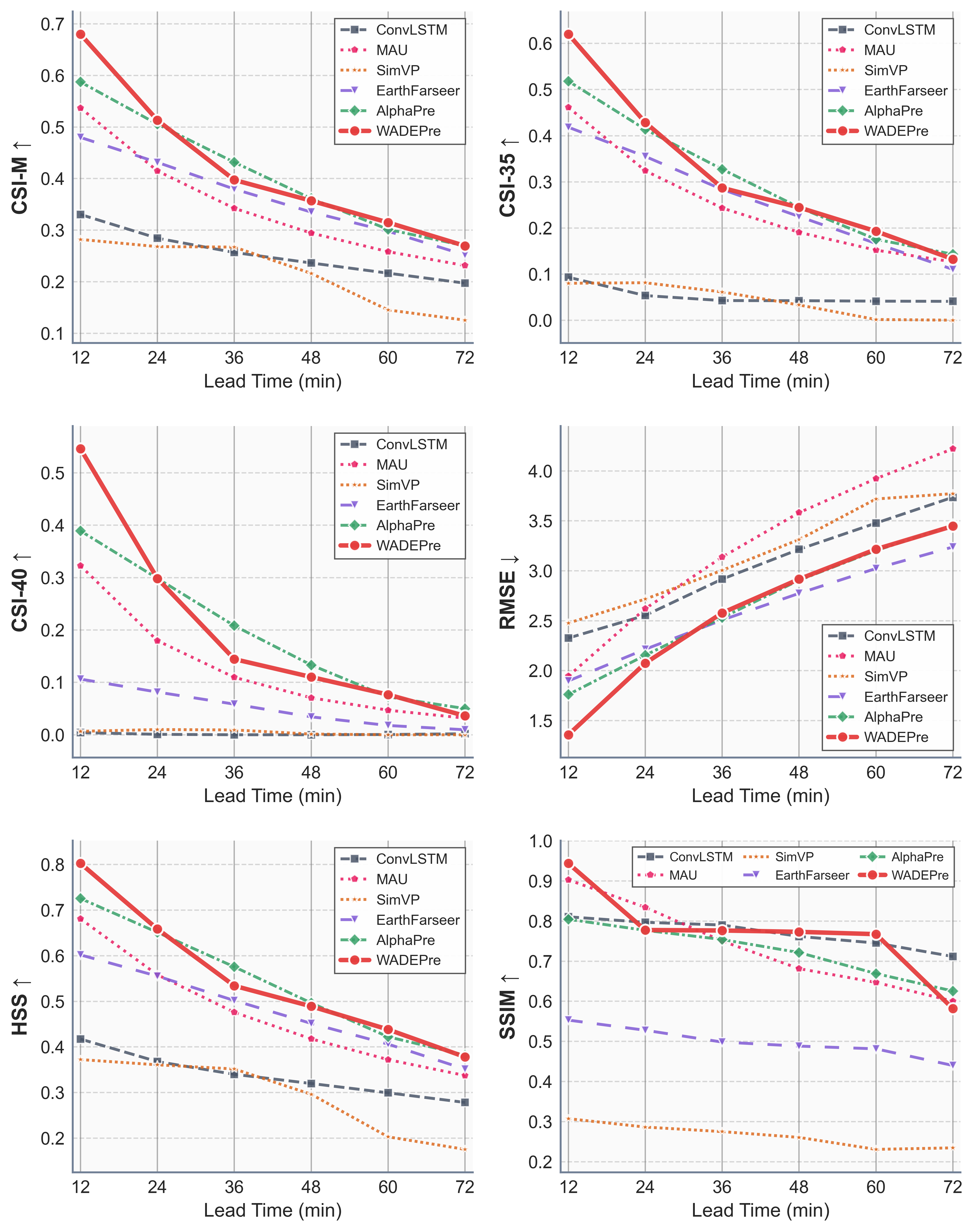}
  \caption{\textbf{Comprehensive temporal evaluation on the Shanghai Radar benchmark}. Performance comparison of all six metrics across a 72-minute lead time. WADEPre (red) achieves a dominant position across all metrics, including the lowest RMSE and highest extreme event capture (CSI-40).}
  \label{fig.metrics_all_shanghai}
\end{figure}

Figure~\ref{fig.metrics_all_shanghai} illustrates the results for the Shanghai Radar dataset. Despite the distinct climatic features and varied intensity thresholds (CSI-35 and CSI-40), our model consistently maintains a substantial advantage. A noticeable decline in performance is observed at the +36 minute lead time, indicating that training parameters may necessitate dataset-specific fine-tuning; nevertheless, WADEPre remains the most effective model overall. As evidenced in the CSI-40 and SSIM subplots, WADEPre markedly surpasses baseline models, demonstrating robust generalization capabilities and operational viability for long-term nowcasting.

\section{Case Study}

To further demonstrate the model's generalization across distinct meteorological regimes, we present three representative case studies: a discrete-cell event associated with \textit{tornado} genesis, a large-scale \textit{heavy rain} event, and a linear \textit{squall line}.

\subsection{Tornado Case: Discrete Cell Preservation}

\begin{figure}[htbp]
  \centering
  \includegraphics[width=\columnwidth]{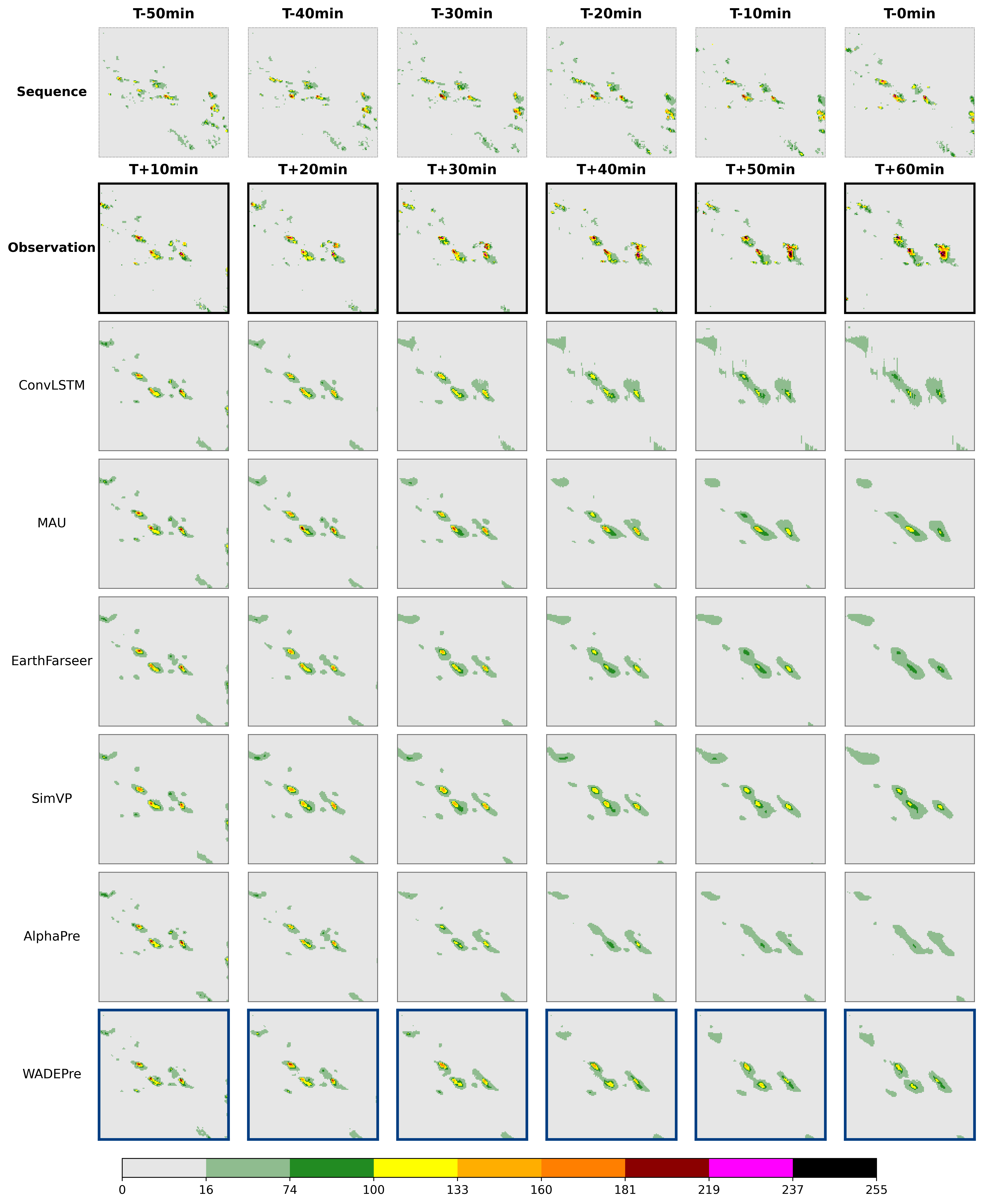}
  \caption{\textbf{Qualitative visualization of a Tornado-genesis event (ID: S843733)}. This case features scattered, high-intensity discrete cells. While baselines blur these isolated structures into the background, WADEPre (blue frames) preserves the sharp, distinct boundaries of the convective cores throughout the 60-minute lead time.}
  \label{fig.case_study_tornado}
\end{figure}

Figure~\ref{fig.case_study_tornado} illustrates a tornado-formation event (ID: S843733), distinguished by dispersed, distinct convective cells with modest yet vigorous cores.

\begin{itemize}
    \item \textbf{Challenge:} The principal challenge resides in maintaining the high-frequency singularities of these isolated cells. As lead time extends, conventional baseline models (for example, ConvLSTM and EarthFarseer) tend to quickly diminish these small objects into faint background noise via pixel-wise averaging.
    \item \textbf{Our Result:} WADEPre maintains topological distinctness at T+60 minutes by utilizing the D-Net to model high-frequency details and accurately delineate sharp convective core boundaries, thereby preventing the ``wash-out'' effect commonly observed in alternative methodologies.
\end{itemize}

\subsection{Heavy Rain Case: Structural Fidelity}

\begin{figure}[htbp]
  \centering
  \includegraphics[width=\columnwidth]{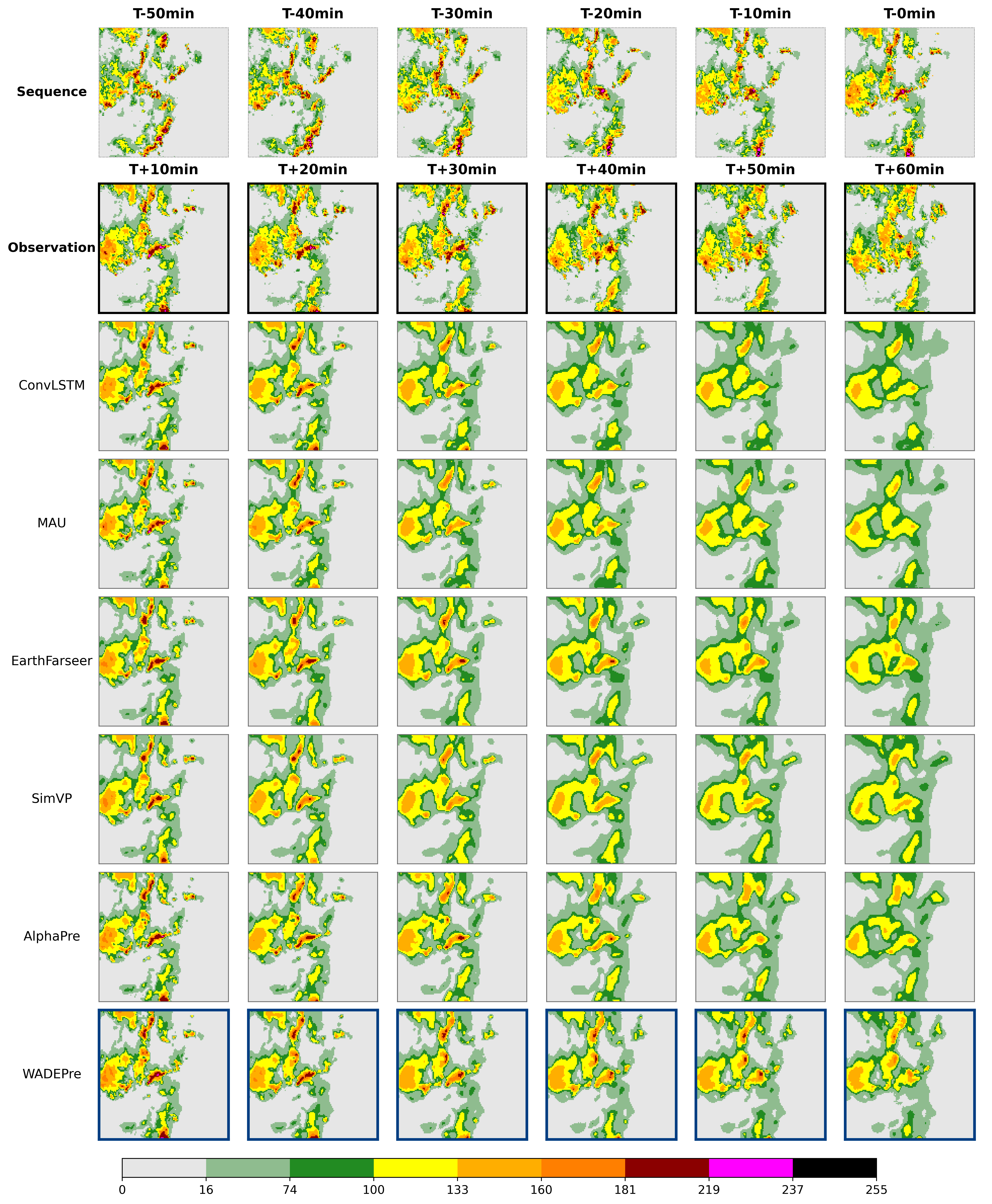}
  \caption{\textbf{Qualitative visualization of a Heavy Rain event (ID: S827031)}. This case involves a large-scale convective system. WADEPre (blue frames) effectively mitigates smoothing, retaining the rich internal texture and high-intensity gradients (red/black regions) that are lost in the baseline predictions.}
  \label{fig.case_study_heavyrain}
\end{figure}

Figure~\ref{fig.case_study_heavyrain} illustrates a significant heavy rain event (ID: S851158), characterized by a complex mesoscale convective system with intricate internal structure.

\begin{itemize}
    \item \textbf{Challenge:} For these extensive systems, the primary challenge lies in preventing the deterioration of internal structural details (texture) into a homogeneous, smooth mass.
    \item \textbf{Our Result:} WADEPre demonstrates superior textural fidelity. Unlike models such as SimVP and MAU, which generate overly smoothed predictions that diminish internal intensity gradients, WADEPre precisely reconstructs the complex spatial hierarchy of storms, preserving fine-grained intensity fluctuations within the primary echo band.
\end{itemize}

\subsection{Squall Line Case: Structural Fidelity}

\begin{figure}[htbp]
  \centering
  \includegraphics[width=\columnwidth]{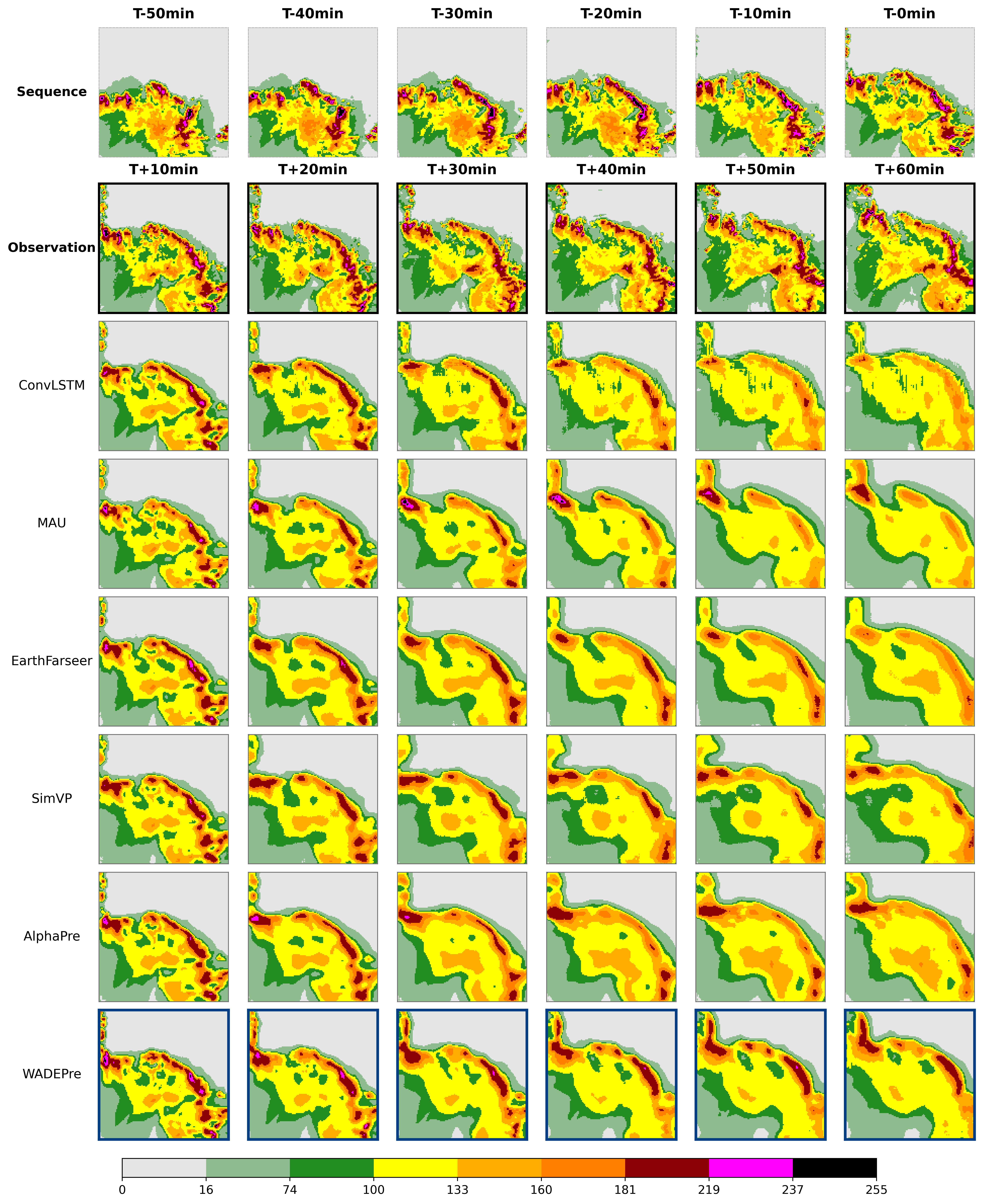}
  \caption{\textbf{Qualitative visualization of a Squall Line event (ID: S825535)}. This case involves a linear convective system with sharp boundaries. WADEPre (blue frames) effectively mitigates spectral smoothing, preserving morphological consistency and high-intensity peaks (pink/black regions) that otherwise dissipate into background noise in baseline predictions.}
  \label{fig.case_study_line}
\end{figure}

Figure~\ref{fig.case_study_line} illustrates the forecast outcomes for a linear squall line event, distinguished by an elongated, high-intensity echo band with well-defined convective boundaries.

\begin{itemize}
    \item \textbf{Challenge:} The primary challenge resides in preserving the linear organization and distinct gradients of the squall line over extended time horizons. Conventional pixel-wise models generally tend to diminish these sharp boundaries beyond $T+30$ minutes, resulting in the organized system deteriorating into a diffuse, structureless cloud.
    \item \textbf{Our Result:} WADEPre demonstrates superior kinematic consistency. By disentangling large-scale advection from local intensity changes, it successfully maintains the morphological integrity of the squall line. It accurately predicts the propagation of the high-intensity leading edge up to $T+60$ minutes.
\end{itemize}

\section{Implementation Details}
\label{appdix.implementation}

Our experiments utilize the PyTorch Lightning framework, with training distributed across four NVIDIA H100 GPUs. The random seed is set to 42 throughout all experiments to ensure consistency. The specific hyperparameters for WADEPre and the training process are summarized below.

\begin{itemize}
	\item \textbf{Wavelet Transform}: The wavelet decomposition level $l$ is set to 3, and we use the bior2.4 wavelet.
    \item \textbf{Approximation Network:} The channel dimension is set to 256, with 3 stacked Spatio-Temporal Blocks (STBlocks).
    \item \textbf{Details Network:} The base feature dimension is 128. The FPN backbone uses channel depths of $[64, 128, 256]$, while the IDR module operates at 64.
    \item \textbf{Refiner:} The hidden dimension size is set to 576.
    \item \textbf{Curriculum Training Strategy:} The loss balancing coefficients are set to $\lambda_{D}=0.05$ and $\lambda_{\text{Mixed}}=0.005$. For the curriculum learning schedule, the dynamic weight decays linearly over $T_\text{decay}=3000$ steps, with a lower bound of $\lambda_{\text{min}}=0.01$.
    \item \textbf{Optimization:} We use the AdamW optimizer with a learning rate of 1.5e-4, betas = (0.9, 0.995), and weight decay = 0.01. Float precision is 32. CosineAnnealingLR with T max set to 200. 
\end{itemize}

\end{document}